\documentclass[aps,onecolumn,preprint,groupedaddress,superscriptaddress,showpacs,nofootinbib]{revtex4-1}

\usepackage{graphicx}
\usepackage{color}
\usepackage{amsmath}
\usepackage{amsfonts}
\usepackage{amssymb}
\usepackage{dcolumn}
\usepackage{hyperref}
\usepackage{enumerate}
\usepackage{bbold}
\usepackage{cancel}
\usepackage{mathtools}
\newcommand{\BEQ}{\begin{equation}}     
\newcommand{\BEA}{\begin{eqnarray}}
\newcommand{\BD}{\begin{displaymath}}
\newcommand{\EEQ}{\end{equation}}       
\newcommand{\EEA}{\end{eqnarray}}
\newcommand{\ED}{\end{displaymath}}

\DeclarePairedDelimiter\ceil{\lceil}{\rceil}

\def\lb#1{{\color{black}{#1}}}

\begin{document}
  \title{Clout, Activists and Budget: The Road to Presidency}
\author{L. B\"{o}ttcher}
\email{lucasb@ethz.ch}
\affiliation{ETH Zurich, Wolfgang-Pauli-Strasse 27, CH-8093 Zurich, Switzerland}
\author{H. J. Herrmann}
\affiliation{ETH Zurich, Wolfgang-Pauli-Strasse 27, CH-8093 Zurich, Switzerland}  
\affiliation{
Departamento de F\'isica, Universidade Federal do Cear\'a, \\
60451-970 Fortaleza, Cear\'a, Brazil}
\author{H. Gersbach}
\affiliation{ETH Zurich, Z\"urichbergstrasse 18, CH-8092 Zurich, Switzerland} 
\date{\today}
\begin{abstract} 
Political campaigns involve, in the simplest case, two competing campaign groups which try to obtain a majority of votes. We propose a novel mathematical framework to study political campaign dynamics on social networks whose constituents are either political activists or persuadable individuals. Activists are convinced and do not change their opinion and they are able to move around in the social network to motivate persuadable individuals to vote according to their opinion. We describe the influence of the complex interplay between the number of activists, political clout, budgets, and campaign costs on the campaign result. We also identify situations where the choice of one campaign group to send a certain number of activists already pre-determines their victory. Moreover, we show that a candidate's advantage in terms of political clout can overcome a substantial budget disadvantage or a lower number of activists, as illustrated by the US presidential election 2016.
\end{abstract}
\maketitle
\section*{Introduction}
Due to the advancement of network science, it has become possible to incorporate the salient features of opinion formation and spreading into large-scale network models \cite{sznajd2000,deffuant2000,hegselmann2002,krapivsky2003,schwaemmle2007,jao2009,ramos15}. The study of political mobilization is one example of such opinion dynamics, which allowed remarkable observations such as the identification of universal features in elections \cite{fortunato07}, the importance of easily persuadable individuals for opinion cascades \cite{watts07}, the impact of online social influence \cite{bond2012} and the necessity of social reinforcement in order to convince people \cite{chwe99,granovetter78,rogers2010diffusion}.

In recent years, political mobilization and the associated campaigns became more and more sophisticated due to online social media, the availability of data on personal preferences and the huge financial campaign support\footnote{Insidegov.com, Compare Presidential Candidates 2016, \url{http://presidential-candidates.insidegov.com/}, retrieved August 31, 2017.} \cite{ramswell2017,boettcher_petitions}. Still, the influence of campaign characteristics such as the number of political activists, political clout or the size of the campaign budget on actual outcomes is not well understood. A more profound understanding of political campaigns is necessary to appropriately interpret the corresponding outcomes and to possibly redesign certain aspects of campaign policies \cite{Gersbach2017}. Thus, we  propose a novel mathematical framework to describe political campaign dynamics on networks. More specifically, we account for the fact that mobile political activists can convince persuadable individuals under certain efficiency and cost restrictions. We demonstrate the existence of one unique stationary solution and rigorously describe the interplay between activists, political clout, budgets and campaign costs on this state, thereby understanding how candidates can win elections. We find that a given budget might allow different choices of the number of activists, such that campaign groups find themselves in an activist-choice game. In these strategic situations, a campaign group chooses activists as a best response to the choice of the other campaign group, which may be crucial for winning the election. Interestingly, some activist combinations lead to strong competition and thus to a large campaign budget. Furthermore, an advantage in terms of political clout can overcome a substantial budget disadvantage or a lower number of activists. This is illustrated for the US presidential election of 2016, in which the winner had a huge budget disadvantage and a lower number of activists to support him.

Election campaigns are an essential component of democratic politics. They have been studied extensively by scholars across social sciences and political sciences, in particular, as surveyed recently in Ref.~\cite{jacobson2015_2}. Three characteristics from this rich body of literature are particularly important for the construction of our model and its investigation:
\begin{enumerate}
\item Some voters are amenable to persuasion and opinion switching, i.e., they might be induced to vote for either candidate if approached by a corresponding campaign activist. In the US presidential election of 2016, the proportion of undecided voters was estimated to be around 20\% to 25\% \cite{kenski2010}. The share of persuadable voters is also large for elections below presidential level,
\item The most effective tactics in campaigns are personal interactions between activists and persuadable voters, such as door-to-door interaction or phone calls \cite{jacobson2015},
\item \lb{Money matters in campaigns. The benefit of campaign expenditures is positive, but typically low and subject to rapid decay \cite{jacobson2015}.}
\end{enumerate}
\section*{Materials and methods}
\begin{figure}
\centering
\includegraphics[width=0.6\textwidth]{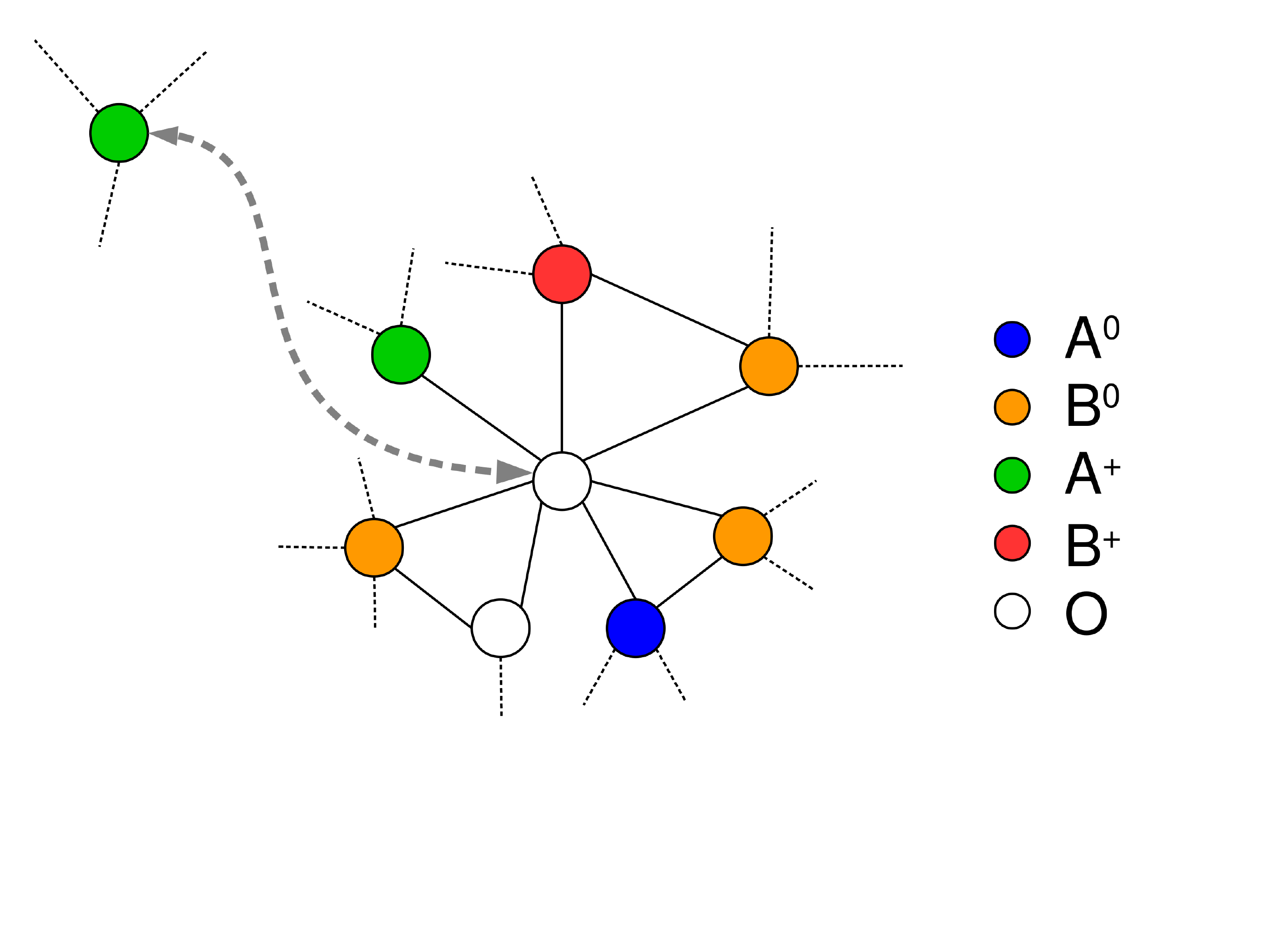}
  \caption{\textbf{Illustration of the campaign model.} An illustration of the campaign model where nodes and their corresponding edges are represented by colored circles and black lines respectively. In this example, an $A^+$ activist node is going to occupy an empty node $O$ to convince  $B^0$ nodes. The utility condition $\rho_A \#B^0-c_A > 0$ in this case is, for example, fulfilled for $\rho_A=1$ and $c_A=2$.}
 \label{fig:model}
\end{figure}
We consider a network with $N$ nodes. Each node is either occupied by an individual having an opinion $A$ or $B$ or it is in an unoccupied state $O$. The total number of nodes is thus $N=N_A+N_B+N_O$. Both campaign groups consist of political activists $A^+$ and $B^+$ and persuadable individuals $A^0$ and $B^0$, such that $N_A=N_{A^+}+N_{A^0}$ and $N_B=N_{B^+}+N_{B^0}$. We illustrate the campaign model in Fig.~\ref{fig:model}. Activists have a fixed opinion and they are able to move from their current location to unoccupied nodes to convince persuadable individuals who do not change their location. Each campaign group is given a finite budget $\mathcal{B}_A$ and $\mathcal{B}_B$ respectively, and a certain cost $c_A$ and $c_B$, respectively, has to be paid for every attempt at convincing a group of persuadable individuals. More specifically, an activist $A^+$ has the possibility to occupy an empty node $O$ from any place in order to change with probability $\rho_A \in (0,1]$ the opinion of neighboring $B^0$ nodes. The activist $A^+$ only moves and convinces other $B^0$ nodes if the utility condition $\rho_A \# B^0 -c_A >0$ is fulfilled and if enough budget $\mathcal{B}_A>0$ is available. Here $\# B^0$ represents the number of neighboring nodes of the empty node which are in state $B^0$. If the $A^+$ activist does move onto the empty node, independent of the number of persuaded $B^0$ nodes, the cost $c_A$ is deducted from the budget $\mathcal{B}_A$ \cite{boettcher14}. The same applies to activists $B^+$, however, with probability $\rho_B$, cost $c_B$ and budget $\mathcal{B}_B$.

\lb{From now on, we consider the fractions $a=A/N$, $b=B/N$ and $o=O/N$ of the states $A$, $B$ and $O$ respectively.} The initial values are given by $a^0(0)$, $b^0(0)$, $a^+(0)$, $b^+(0)$ and $o(0)$.
By definition of our model, we know that only $a^0$ and $b^0$ change over time. Thus, $\dot{a}^+=0$, $\dot{b}^+=0$, $\dot{o}=0$ and thus $a^+(t)=a^+(0)$, $b^+(t)=b^+(0)$ and $o(t)=o(0)$. It is now possible to derive mean-field rate equations of $\dot{a}^0(t)$ and $\dot{b}^0(t)$ by assuming a perfectly mixed population in the thermodynamic limit. Mean-field approximations have been proven useful to gain important insights about a given spreading dynamics \cite{keeling-rohani2008,marro05}, and many real-world social networks are well described by mean-field approximations \cite{gleeson12}. Our analytical results are supported by stochastic kinetic Monte Carlo simulations \cite{gillespie76,gillespie77}. For analytical tractability, we assume a \emph{regular network with fixed degree} $k$, i.e.~a fixed number of neighbors. A mean-field approach for general degree distributions $f_k$ is presented in Refs.~\cite{boettcher162, boettcher171} and is based on an additional weighted sum accounting for different degrees in the network. We only focus on the derivation of $\dot{a}^0(t)$, since $b^0=1-a^+-b^+-o-a^0$. Our first observation is that the utility condition $ \rho_A \# B^0 -c_A >0$ corresponds to a threshold \cite{granovetter78,chwe99,watts02,macy07,easley2010,gleeson2013,boettcher162,boettcher171} in the sense that $A^+$ activists only move to empty nodes whose neighborhoods contain at least $\ceil*{\frac{c_A}{\rho_A}}$ persuadable nodes in state $B^0$. Here $\ceil*{\cdot}$ denotes the ceil function since the quantity $\frac{c_A}{\rho_A}$ might not have an integer value. We define the transition functions $f_{A^0\rightarrow B^0}(t)$ and $f_{B^0\rightarrow A^0}(t)$ describing transitions $A^0\rightarrow B^0$ and $B^0\rightarrow A^0$, respectively, and find for our dynamics
\begin{equation}
\begin{split}
\dot{a}^0(t)= & f_{B^0\rightarrow A^0}(t)-f_{A^0\rightarrow B^0}(t)  \\
= & o \rho_A \Theta(\mathcal{B}_A) \sum_{j=\ceil*{\frac{c_A}{ \rho_A}}}^{k} \sum_{l,m.n,r \in S_{k- j}} j \binom{k}{j,l,m,n,r} {b^0}^j {a^+}^{l+1} {b^+}^m {a^0}^n o^r\\ 
& -o \rho_B \Theta(\mathcal{B}_B) \sum_{j=\ceil*{\frac{c_B}{ \rho_B}}}^{k} \sum_{l,m.n,r \in S_{k -j}} j \binom{k}{j,l,m,n,r}{a^0}^j {a^+}^l {b^+}^{m+1} {b^0}^n o^r,
\end{split}
\label{eq:rate}
\end{equation}
where $\Theta(\cdot)$ denotes the Heaviside step function accounting for the fact that opinion changes only occur when enough budget is available and $\binom{k}{j,l,m,n,r}$ is the multinomial coefficient, such that $j+l+m+n+r=k$ and $S_{k- j}=\left\{ l,m,n,r \geq 0, l+m+n+r=k-j \right\}$. The prefactor $o$ in Eq.~\eqref{eq:rate} accounts for the necessity for a node to be empty before an activist can occupy it, and $\rho_A$, $\rho_B$ denote the probabilities to persuade others, henceforth \emph{convincing probabilities}. The first double sum describes transitions $B^0\rightarrow A^0$ where at least $\ceil*{\frac{c_A}{\rho_A}}$ nodes of type $B^0$ need to be available according to the utility condition. The second term describes reverse transitions $A^0\rightarrow B^0$. We notice that \emph{no change} occurs if $\ceil*{\frac{c_A}{\rho_A}}>k$ and $\ceil*{\frac{c_B}{\rho_B}}>k$. The cost deductions from the respective budgets are described by 
\begin{align}
& \dot{\mathcal{B}}_A(t)=-\frac{c_A}{\rho_A} f_{B^0\rightarrow A^0}(t), \label{eq:budget_A} \\
& \dot{\mathcal{B}}_B(t)=-\frac{c_B}{\rho_B} f_{A^0\rightarrow B^0}(t),
\label{eq:budget_B}
\end{align}
with the initial conditions $\mathcal{B}_A(0)=\mathcal{B}_A^0$, $\mathcal{B}_B(0)=\mathcal{B}_B^0$ where $\mathcal{B}_A^0$ and $\mathcal{B}_B^0$ are the campaign budgets of the two campaign groups, respectively. \lb{In general, our model could also be extended to more than two campaign groups, i.e.~more than two candidates. It is then necessary to account for the additional corresponding costs, convincing probabilities and transition functions.}

In the \emph{Appendix}, we demonstrate that the dynamics captured by Eq.~\eqref{eq:rate} exhibits a unique stationary solution for a given set of initial conditions. Furthermore, we show that there is a complex interplay between the fractions of activists, convincing probabilities and costs. In the case of $\rho_A=\rho_B$ and $c_A=c_B$, the campaign group with the larger fraction of activists obtains the majority. Otherwise, different convincing probabilities and different costs lead to a shift of the stationary states. In particular, a large convincing probability and small cost is advantageous for the respective campaign group. Convincing probabilities cannot be interchanged with the fractions of activists. \lb{Small convincing probabilities and large costs lead to longer transients and to stationary states that are closer together, i.e., small gaps between the final campaign results. Unlike the activist fractions, convincing probabilities and costs also enter in the threshold condition.} We also discuss one example in the \emph{Appendix}, which illustrates how the influence of $\rho$ and $c$ might, surprisingly, lead to a change in the majority structure.
\section*{Results}
\label{sec:results}
\subsection*{Budget limitations}
\begin{figure}
\centering
\includegraphics[width=\textwidth]{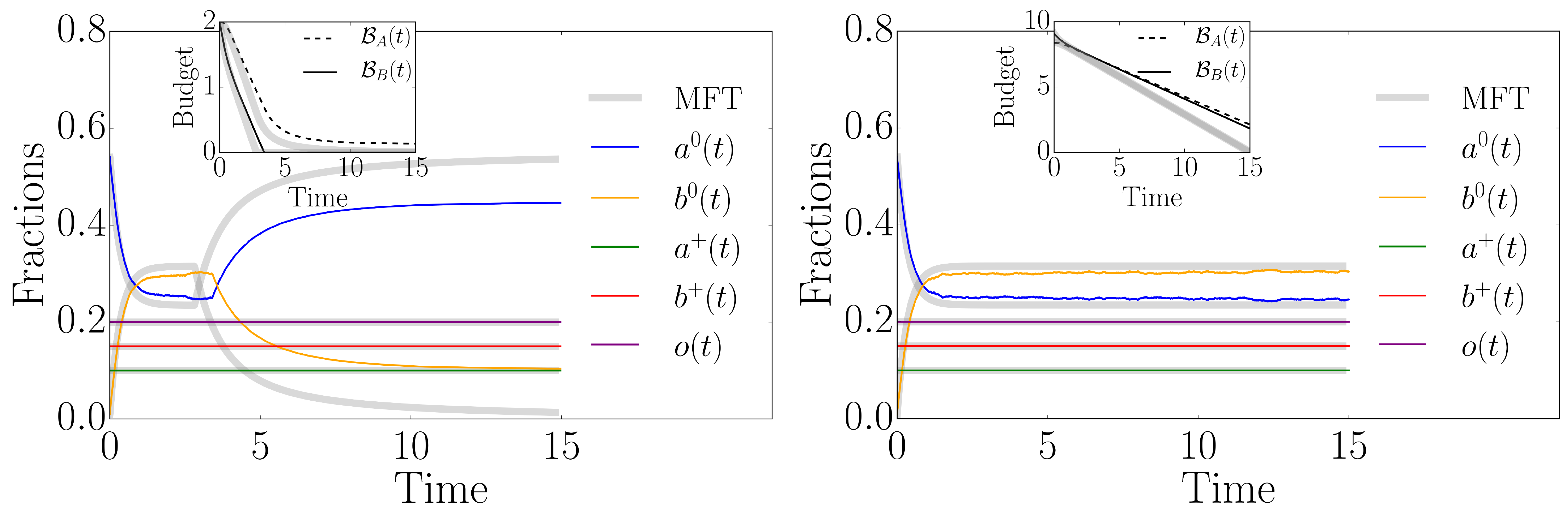}
  \caption{\textbf{Influence of budget restrictions.} The time evolution of the fractions of persuadable nodes $a^0(t)$, $b^0(t)$, activists $a^+(t)$, $b^+(t)$, empty nodes $o(t)$ as well as the budgets $\mathcal{B}_A(t)$ and $\mathcal{B}_B(t)$. The shown curves are the results of a stochastic simulation on a random regular network with $N=10^5$ nodes, $k=10$ and the following initial conditions: $a^0(0)=0.54$, $b^0(0)=0.01$, $a^+(0)=0.1$, $b^+(0)=0.15$ and $o(0)=0.2$. The convincing probabilities are fixed to $\rho_A=\rho_B=1$ and the costs to $c_A=c_B=2$. The thick grey curves are the numerical mean-field solutions of Eq.~\eqref{fig:stochastic_1}. In the left panel, the budgets are exhausted after some time due to the small initial values $\mathcal{B}_A(0)=2$ and $\mathcal{B}_B(0)=2$, and thus the evolution changes. On the other hand, in the right panel, the initial budgets $\mathcal{B}_A(0)=8.4$ and $\mathcal{B}_B(0)=9.1$ are chosen to be large enough not to influence the evolution during the given time horizon.}
 \label{fig:stochastic_1}
\end{figure}
A political campaign typically lasts a fixed time $T$. We assume that political campaigns have information about the convincing probabilities $\rho_A$ and $\rho_B$ as well as the costs $c_A$ and $c_B$. A typical goal of each campaign is to convince as many people as possible by spending the available budget in such a way that there is no money left at time $T$ when campaigns start at time 0. Intuitively, the amount of budget spent does not only depend on the costs and the convincing probabilities but also on the number of activists in each campaign group. In particular, the budget $\mathcal{B}_B$ will be exhausted faster than budget $\mathcal{B}_A$ if the initial budgets, the costs and the convincing probabilities are equal but the fraction of activists of campaign group $B$ is larger than the one of group $A$. This might be problematic if the campaign lasts longer than the time until budget exhaustion, since campaign group $B$ will lose its majority sooner or later. We illustrate this effect in Fig.~\ref{fig:stochastic_1} (left) where numerical simulations are presented together with the corresponding mean-field approximation. The finite size and the finite degree of the underlying random regular network lead to deviations from the mean-field perfect mixing assumption. For large degrees one finds better agreement between simulations and mean-field results \cite{boettcher162}. Still, the simulated dynamics is qualitatively captured by Eq.~\eqref{eq:rate}. To better understand the influence of budget restrictions, we now focus on two relevant points related to the latter example: (i) we want to determine the minimum necessary budget for given fractions of activists $a^+$, $b^+$ and a given campaign duration, and (ii) we want to discuss strategies \emph{when} to send \emph{how} many activists to effectively use a given budget so that it lasts until the campaign ends.

We begin with the discussion of point (i) and find that the minimum necessary budgets are 
\begin{align}
& \mathcal{B}_A^{\mathrm{min}}(T)=\int_0^T \frac{c_A}{\rho_A}f_{B^0\rightarrow A^0}(t) \, \mathrm{d}t, \label{eq:min_budget_A} \\
& \mathcal{B}_B^{\mathrm{min}}(T)=\int_0^T \frac{c_B}{\rho_B}f_{A^0\rightarrow B^0}(t) \, \mathrm{d}t,
\label{eq:min_budget_B}
\end{align}
according to Eqs.~\eqref{eq:budget_A} and \eqref{eq:budget_B}.
To solve the integrals defined by Eqs.~\eqref{eq:min_budget_A} and \eqref{eq:min_budget_B}, one first has to solve Eq.~\eqref{eq:rate}, assuming sufficient resources, i.e.~$\Theta(\mathcal{B}_A)=1$ and $\Theta(\mathcal{B}_B)=1$ for all times. This solution has then to be used to compute the integrals, which yields unique solutions. We again look at Fig.~\ref{fig:stochastic_1} (left), where the initial budget $\mathcal{B}_B(0)=2$ turns out to be insufficient for a duration $T=15$. What is the necessary minimum in this case? We solve Eqs.~\eqref{eq:min_budget_A} and \eqref{eq:min_budget_B} and find that the mean-field initial budgets are $\mathcal{B}_A^{\mathrm{min}}(T)=8.4$ and $\mathcal{B}_B^{\mathrm{min}}(T)=9.1$. We clearly see in Fig.~\ref{fig:stochastic_1} (right) that these budgets are sufficient. Due to the perfect mixing assumption, the mean-field dynamics tends to be slightly faster than the one of the simulations.

\subsection*{Strategic choices of activists}
\begin{figure}
\centering
\includegraphics[width=\textwidth]{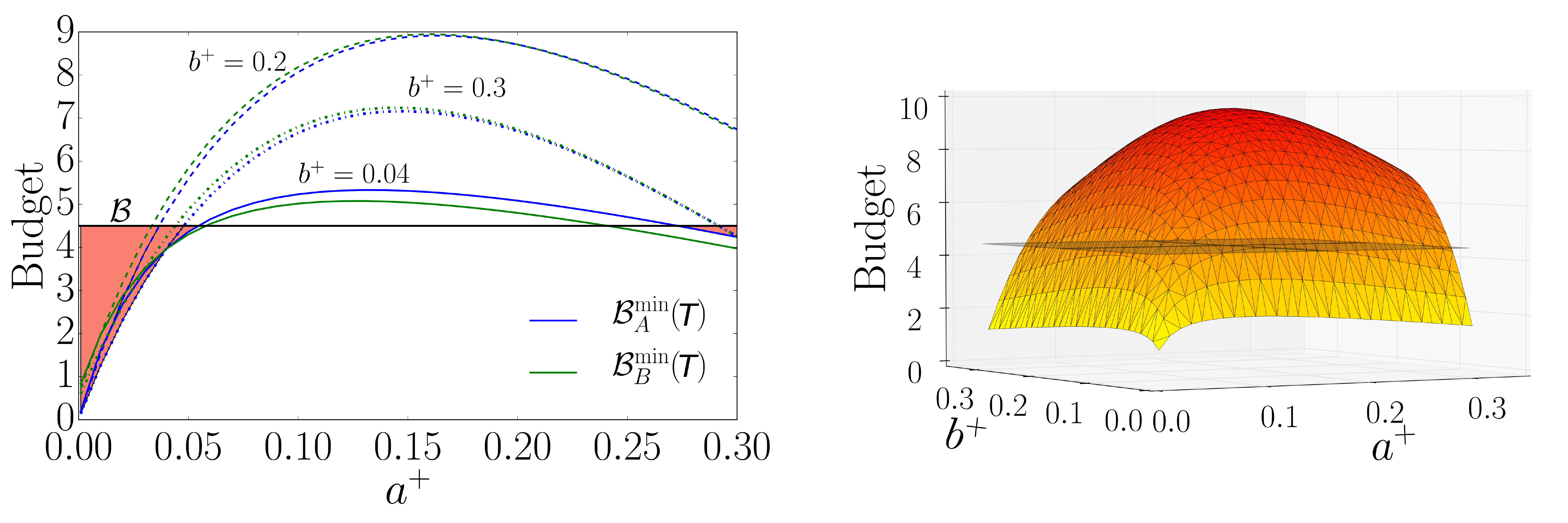}
  \caption{\textbf{Minimal budgets for different activist fractions.} We set $k=10$ and used the following initial conditions: $a^0(0)=b^0(0)=(1-a^+-b^+-o)/2$ with $o(0)=0.2$. The convincing probabilities are fixed to $\rho_A=\rho_B=1$ and the costs to $c_A=c_B=2$. In the left panel, the minimal budgets $\mathcal{B}_A^{\mathrm{min}}(T)$ and $\mathcal{B}_B^{\mathrm{min}}(T)$ according to Eqs.~\eqref{eq:min_budget_A} and \eqref{eq:min_budget_B} are displayed as a function of the fraction $a^+$ for different values of $b^+$. The black solid line indicates a given budget $\mathcal{B}=4.5$ for campaign of duration $T=15$ and the orange areas represent the corresponding regimes of possible choices for $a^+$ for which the given budget $\mathcal{B}$ suffices. The right panel shows a three-dimensional illustration of the budget $\mathcal{B}_A$ as a function of $a^+$ and $b^+$. The plane again represents a given budget $\mathcal{B}=4.5$. All values of $a^+$ and $b^+$ leading to budget values below or on the plane are possible choices to have a sufficient budget.}
 \label{fig:budget_1}
\end{figure}
For the discussion of point (ii), it is important to realize that while Eqs.~\eqref{eq:min_budget_A} and \eqref{eq:min_budget_B} yield unique minimal budgets for given $a^+$, $b^+$ and $T$, finding the two fractions of activists, for given campaign budgets and a duration $T$, is not a unique operation.

Here we solely focus on choosing constant fractions of activists $a^+,b^+$ within the interval $ (0,0.3]$, since $a^+$ and $b^+$ are typically not too large. Fig.~\ref{fig:budget_1} (left) helps to gain some intuition for the existence of multiple possible choices of $a^+$ and $b^+$ for given campaign budgets. We assumed equal convincing probabilities and costs. In this case, the campaign group with the larger fraction of activists wins the campaign in the limit of long times, as shown in the \emph{Appendix}. Fig.~\ref{fig:budget_1} (left) shows the solutions of Eqs.~\eqref{eq:min_budget_A} and \eqref{eq:min_budget_B} for $a^+\in(0,0.3]$ and three values of $b^+\in\{0.04, 0.2, 0.3\}$. All solutions are parabolas with a characteristic maximum originating from the competition between the two campaign groups. If one or the other campaign dominates, a relatively small budget suffices due to the small average number of opinion changes per unit of time, i.e.~small transition functions in Eqs.~\eqref{eq:budget_A} and \eqref{eq:budget_B}. However, if the two activist fractions are of similar size, the resulting competition leads to a larger necessary budget due to the larger transition functions in Eqs.~\eqref{eq:budget_A} and \eqref{eq:budget_B}. Considering a limited small budget, it is thus unfavorable for campaign groups to have activist numbers facilitating strong competition. Consequently, choosing the right number of activists is crucial for winning the election. In the more general case, one has to take the other parameters $\rho_A$, $\rho_B$, $c_A$ and $c_B$ into account which can lead to shifted maxima.

We now want to graphically determine a tuple $(a^+,b^+)$ such that a budget $\mathcal{B}$ given to each campaign group suffices throughout the campaign duration. We assume a budget $\mathcal{B}=4.5$ for campaign of duration $T=15$ and observe that, for example, in the case of $b^+=0.04$ it would allow for two solution branches, indicated by the orange areas in Fig.~\ref{fig:budget_1} (left), where $a^+>0.27$ or $a^+<0.05$. In general, a given small budget restricts the possible choices of $a^+$ and $b^+$ to values on or below a plane, as shown in Fig.~\ref{fig:budget_1} (right). If $\mathcal{B}$ exceeds the maximum of the parabolas, all constellations of $a^+$ and $b^+$ are possible for the budgets to be sufficient, i.e.~$a^+, b^+ \in (0,1)$. Interestingly, in the case of a small budget leading to two solution branches as in Fig.~\ref{fig:budget_1} (left), one can interpret the situation as follows. Campaign group $B$, for example, takes a value of $b^+=0.3$. Then group $A$ can either send a very small fraction or one close or equal to $b^+$. In this case, group $B$ would never lose, but depending on the choice of group $A$, the difference between the final numbers of individuals in each campaign group would be smaller for larger $a^+$. However, if group $B$ takes a smaller value such as $b^+=0.04$, campaign group $A$ has the possibility to send a larger fraction to win the campaign. Therefore, the decision for a certain activist fraction made by one campaign group affects the possible choices of the other group. The right initial choice is important for winning the election.

Next we look at the Nash equilibria of the strategic choices in the ensuing activist game as illustrated in Fig.~\ref{fig:budget_1} (left). Formally, we investigate the best responses of the game defined by 
\begin{equation}
\max_{\mathcal{B}_A^{\mathrm{min}}(T)\leq 4.5,~b^+~\mathrm{given}}\{a^+\},~\max_{\mathcal{B}_B^{\mathrm{min}}(T)\leq 4.5,~a^+~\mathrm{given}}\{b^+\}.
\end{equation}
A Nash equilibrium is a constellation $(a^+,b^+)$ that solves both problems. In our example a larger fraction of activists increases the chance of winning and we obtain three best responses for a fixed $b^+\in \{0.04,0.2,0.3\}$
\begin{equation*}
\begin{split}
& (a^+, b^+)_{b^+~\mathrm{given}}=(0.3,0.04), \\
& (a^+, b^+)_{b^+~\mathrm{given}}=(0.03,0.2), \\
& (a^+, b^+)_{b^+~\mathrm{given}}=(0.3,0.3).
\end{split}
\end{equation*}
As a consequence of the parameters and initial conditions used in Fig.~\ref{fig:budget_1}, the tuples obey the following symmetry: $(x, y)_{b^+~\mathrm{given}}=(y,x)_{a^+~\mathrm{given}}$. The game admits multiple equilibria. Besides $(0.3,0.3)$, for instance, there also exists another symmetric equilibrium around $(0.044, 0.044)$. Moreover, two campaigns with identical parameters, clout and budget are compatible with a constellation in which $A$ wins with a majority. The opposite constellation also exists, however. In practice, when the activists are not chosen simultaneously, we can look at how a campaign group can generate a first-mover advantage by choosing a large but still affordable number of activists that forces the other group to choose a lower number of activists. The preceding observation suggests that in a race of two campaigns having identical parameters and budgets, the campaign wins that first has assembled the right number of activists.
\subsection*{Can clout compensate a lower budget and fewer activists?}
\begin{figure}
\centering
\includegraphics[width=\textwidth]{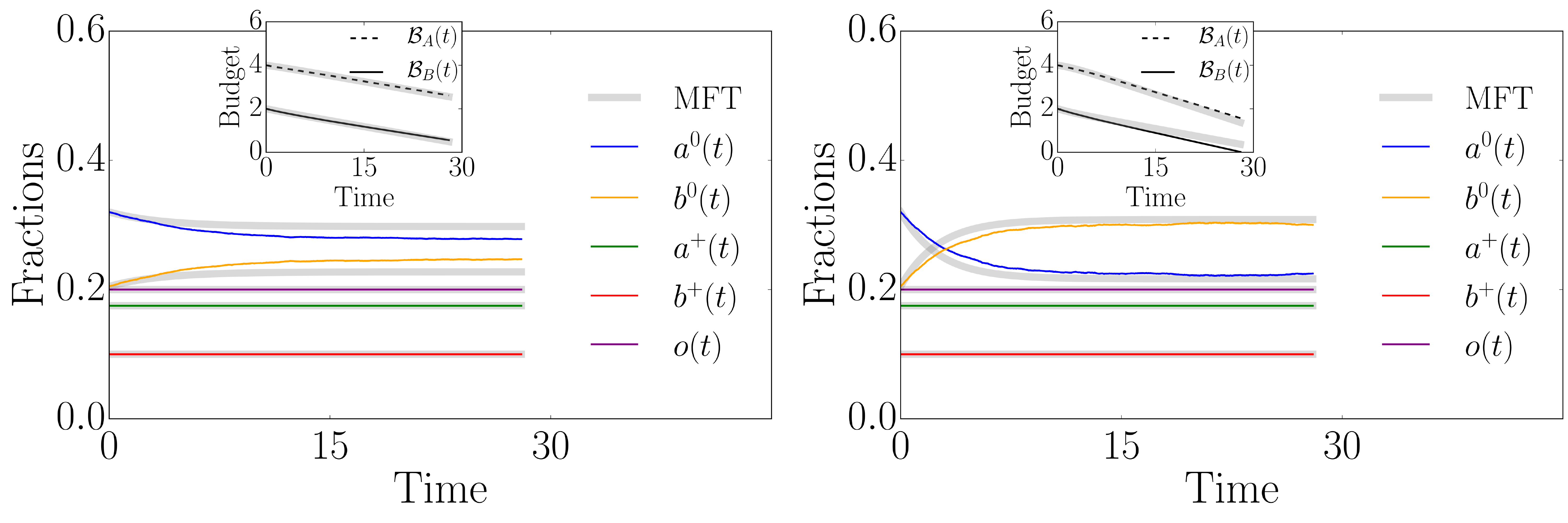}
  \caption{\textbf{Clout versus budget and activists.} The time evolution of the fractions of persuadable voters $a^0(t)$, $b^0(t)$, activists $a^+(t)$, $b^+(t)$, empty nodes $o(t)$ as well as the budgets $\mathcal{B}_A(t)$ and $\mathcal{B}_B(t)$. The shown curves are the results of a stochastic simulation on a random regular network with $N=10^5$ nodes, $k=10$ and the following initial conditions: $a^0(0)=0.32$, $b^0(0)=0.205$, $a^+(0)=0.175$, $b^+(0)=0.1$ and $o(0)=0.2$. The cost is set to $c_A=c_B=0.3$. The grey curves are the numerical mean-field solutions of Eq.~\eqref{fig:stochastic_1}. The left panel shows the situation where both campaign groups have activists with $\rho_A=\rho_B=0.1$. In the right panel, the convincing probability in campaign group $B$ has been increased to a value of $\rho_B=0.18$, leading to a change of the majority structure.}
\label{fig:stochastic_2}
\end{figure}
The model outlined above can be used to study political campaigns for specific parameter environments and to address campaign regulation. As an example we illustrate an application of the model and illustrate a phenomenon observed in the recent presidential election in the US battleground states. Three characteristics are widely discussed. First, Donald Trump had a significantly lower budget than Hillary Clinton \footnote{Insidegov.com, Compare Presidential Candidates 2016, \url{http://presidential-candidates.insidegov.com/}, retrieved August 31, 2017.}. More precisely, he had half of Clinton's budget (404 versus 807 million US dollars). Second, Clinton had a higher number of volunteers than Trump (7\% versus 4\%)\footnote{Ariel Edwards-Levy, Volunteering For A Campaign Or Going To Rallies? You're In The Minority, Huffpost, 11.01.2016, \url{http://www.huffingtonpost.com/entry/campaign-volunteering-rallies-poll_us_581906b8e4b0f96eba968ca7}, retrieved August 31, 2017.}. Third, Trump was able to flip millions of white Obama supporters to his side. We thus investigate whether an advantage in terms of clout could compensate both a significantly lower budget and a lower share of activists. The clout is quantified through the convincing probabilities $\rho_A$ and $\rho_B$. In Fig.~\ref{fig:stochastic_2} we illustrate a campaign situation roughly comparable to the one of the US presidential election 2016. We interpret campaign group $A$ as the Clinton group, whereas group $B$ represents Trump's campaign. We set the initial values of persuadable individuals to $a^0(0)=0.32$ and $b^0(0)=0.205$, taking into account the initial majority structure of the actual US presidential election 2016 poll data\footnote{Poll Chart 2016 General Election: Trump vs.~Clinton, Huffpost Pollster, \url{http://elections.huffingtonpost.com/pollster/2016-general-election-trump-vs-clinton}, retrieved August 31, 2017.}. Considering the aforementioned presidential election characteristics, we set $a^+=0.175$, $b^+=0.1$, $\mathcal{B}_A^0=4$, $\mathcal{B}_B^0=2$ to account for the different budgets and numbers of activists ($\mathcal{B}_A^0/\mathcal{B}_B^0 \approx 807/404$ and $a^+/b^+=7/4$). Furthermore, we assume equal costs $c_A=c_B=0.3$ and first study the case of equal convincing probabilities $\rho_A=\rho_B=0.1$, as illustrated in Fig.~\ref{fig:stochastic_2} (left). \lb{As demonstrated in the \emph{Appendix}, only the relative fraction $\lambda=a^+/b^+$ matters in determining the winning campaign if the convincing probabilities and costs are equal---campaign group $A$ wins if $\lambda > 1$. This is the reason for our choice of $a^+/b^+ = 0.175/0.1=7/4$.} Clearly, the Clinton campaign group $A$ dominates the dynamics due to the larger fraction of activists. However, in Fig.~\ref{fig:stochastic_2} (right), we show that additional clout can compensate a significantly lower number of activists and a lower budget. 
\begin{figure}
\centering
\includegraphics[width=0.45\textwidth]{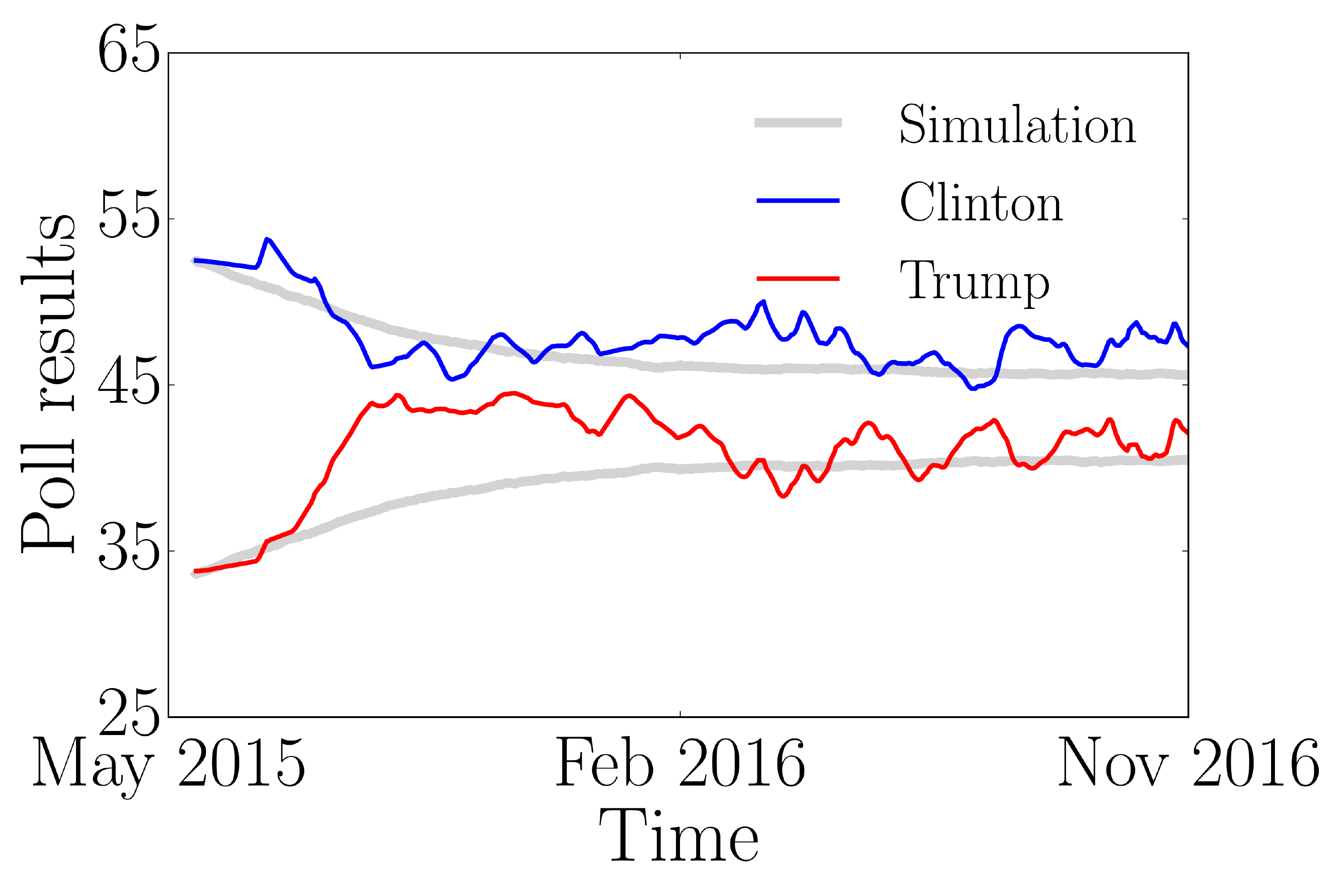}
 \caption{\textbf{Poll and simulation results of the US presidential election 2016.} \lb{Blue and red curves represent the poll results for Hillary Clinton and Donald Trump, respectively. We compare the poll results with our simulation (grey line) using a random regular network with $N=10^5$ nodes, $k=10$ and the following initial conditions: $a^0(0)=0.32$, $b^0(0)=0.205$, $a^+(0)=0.175$, $b^+(0)=0.1$ and $o(0)=0.2$. The cost is set to $c_A=c_B=0.3$ and the convincing probabilities to $\rho_A=\rho_B=0.1$. The results of the simulations have been rescaled to match the initial values of the polls and the corresponding duration. The data has been taken from \url{http://elections.huffingtonpost.com/pollster/2016-general-election-trump-vs-clinton}.}}
\label{fig:trump_vs_clinton}
\end{figure}
\lb{Of course, one cannot say that Donald J.\ Trump won the US presidential elections only because of more political clout. Other factors such as targeted campaigns or the sophisticated use of social media have also been important for his success \cite{trump_media,trump_cambridge}. The example discussed here just illustrates the strong influence of political clout on the campaign outcome.}

This observation has broader implications for the resilience of democracy, as it is related to recent phenomena associated with populism. Candidates who promise to protect citizens from unemployment or income shocks caused by disruptions due to globalization or automation may be able to generate sufficient clout \cite{rodrik17}. Such candidates may win election campaigns even if they will not be able to keep their promises or even enact detrimental policies.

\lb{Finally, we compare the simulation results presented in Fig.~\ref{fig:stochastic_2} (left) with the US presidential election polls 2016, bearing in mind that the poll data did not reproduce the final result, as illustrated in Fig.~\ref{fig:trump_vs_clinton}. For this comparison we rescaled the simulation results to match the initial values of the polls and the corresponding poll duration. As outlined above, our simulations take into account the empirically determined fractions of activists and assume values of $\rho_A=\rho_B=0.1$ and $c_A=c_B=0.3$. It has not been possible to determine these values empirically. Since convincing probabilities and costs are equal in this example, Clinton's campaign group wins due to the larger number of activists. We find that the simulated poll trajectories qualitatively correspond to the actual poll data. Assuming other values of convincing probabilities and costs would imply a different characteristic convergence time scale and a different gap between the final campaign results as shown in the \emph{Appendix}. One would, however, still observe a qualitatively similar convergence behavior.}
\section*{Discussion}
We have developed a novel framework to study political campaign dynamics on social networks based on three essential features: moving activists, political clout and campaign budgets. We illustrated how the complex interplay between these factors determines the success of a campaign group. Our results imply that the right initial choice of the number of activists might lead to an important advantage in winning a campaign. In addition, we also showed that typical campaign characteristics can be taken into account by our model allowing to apply it to the US presidential election of 2016. What is more, our model opens up many further applications. For instance, we could introduce caps on campaign budgets to study how such regulations impact the outcome of elections. Furthermore, our framework allows to integrate additional channels such as media or targeting \cite{boettcher_targeted} which strongly influence campaigns. Finally, by introducing preferences of citizens about policy-making once a candidate is in office, we can identify conditions under which a politician may win and enacts policies that will harm a majority of citizens.
\section*{Acknowledgments}
We acknowledge financial support from the ETH Risk Center and ERC Advanced grant number FP7-319968 FlowCCS of the European Research Council. We thank Moritz Hoferer for his input regarding the strategic choices of activists and for proofreading the manuscript. L.B.~wants to thank Sergio Sol\'{o}rzano for interesting discussions.

\section*{Author Contributions}
\label{sec:contr}
Conceived and designed the experiments: LB HJH HG. Performed the experiments: LB. Analyzed the data: LB. Wrote the paper: LB HJH HG.

\appendix*
\section{Model properties}
\label{sec:properties}
\subsection{Uniqueness of the stationary solution}
\label{sec:uniqueness}
\begin{figure}
\begin{minipage}{0.45\textwidth}
\centering
\includegraphics[width=\textwidth]{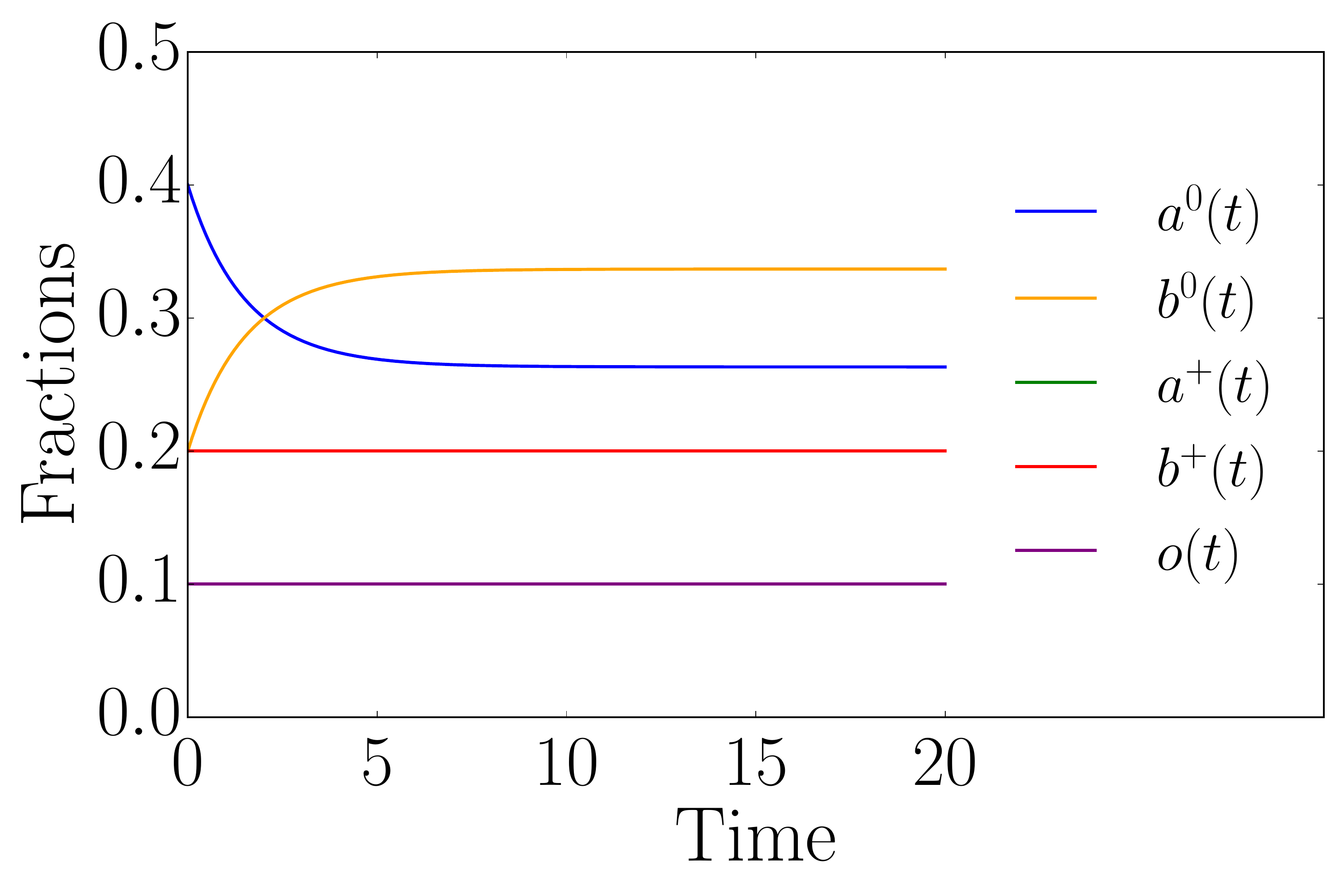}
\end{minipage}
\begin{minipage}{0.45\textwidth}
\centering
\includegraphics[width=\textwidth]{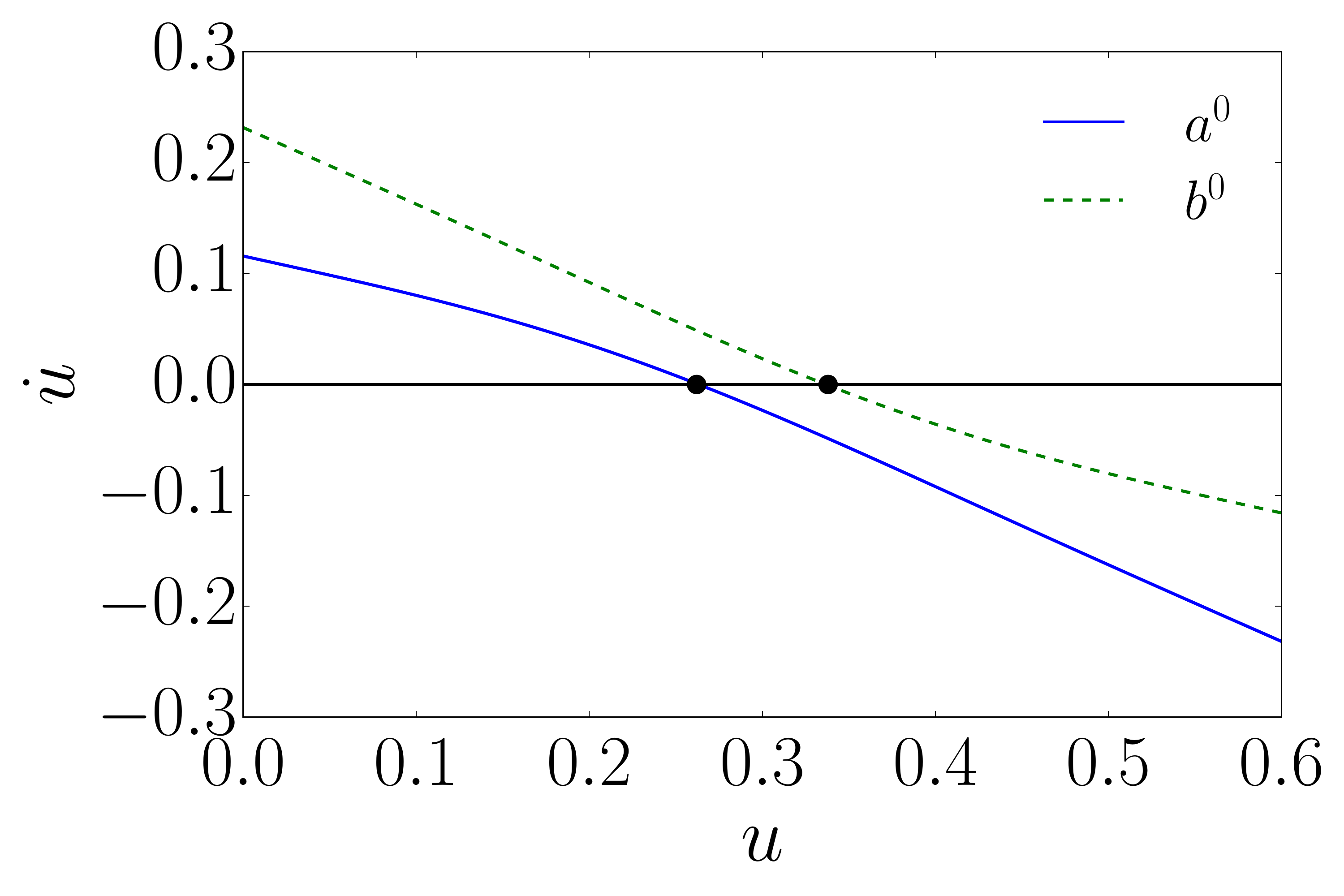}
\end{minipage}
  \caption{\textbf{Time evolution and phase portrait of the campaign model.} The shown curves are a numerical solution of Eq.~\eqref{eq:rate} for $k=10$ with initial conditions $a^0(0)=0.4$, $b^0(0)=0.2$, $a^+(0)=0.1$, $b^+(0)=0.2$ and $o(0)=0.1$. The convincing probability is set to $\rho_A=\rho_B=0.5$ and the costs are $c_A=c_B=2$. The left panel displays the time evolution of the fractions of persuadable nodes $a^0(t)$, $b^0(t)$ and activists $a^+(t)$, $b^+(t)$ as well as empty nodes $o(t)$. In the right panel, the corresponding phase portrait is shown. Black dots represent stable fixed points.} 
 \label{fig:xdotx}
\end{figure}
If not noted otherwise, we assume unlimited resources ($\Theta(\mathcal{B}_A)=1$ and $\Theta(\mathcal{B}_B)=1$ for all times $t$) in the subsequent sections to characterize the stationary states of Eq.~\eqref{eq:rate}. In the case of limited resources, the activist dynamics would just stop in one or the other campaign group. Is there always a unique stationary solution of Eq.~\eqref{eq:rate}, i.e.~a unique stationary $a^0_{\mathrm{st}}$ and $b^0_{\mathrm{st}}$ in the interval $\left(0,a^0(0)+b^0(0)\right]$? We remember that the fractions of activists $a^+$, $b^+$ and of empty nodes $o$ are constant over time. Thus, the dynamics only applies to persuadable nodes leading to trajectories where $a^0(t),~b^0(t)\in \left(0,a^0(0)+b^0(0)\right]$. In Fig.~\ref{fig:xdotx} (left) we show a typical time evolution of our dynamics. The corresponding phase portrait is illustrated in Fig.~\ref{fig:xdotx} (right). We clearly see that there is only one stable fixed point for $a^0$ and $b^0$.

To obtain some insights into the general behavior of the stationary solutions, we simplify Eq.~\eqref{eq:rate} by solely focusing on the dynamical part of the network. More specifically, we introduce an effective degree $k_{\mathrm{eff}}=\ceil*{k (1-a^+-b^+-o)}$ and an effective threshold $m_{\mathrm{eff}}=\ceil*{\frac{c}{\rho} (1-a^+-b^+-o)}$, assuming $\rho_A=\rho_B=\rho$ and $c_A=c_B=c$. Approximating Eq.~\eqref{eq:rate} by introducing an effective degree $k_{\mathrm{eff}}$ means to distribute the fractions $a^0$ and $b^0$ of nodes in a network with degree $k_{\mathrm{eff}}$. An effective threshold $m_{\mathrm{eff}}$ can be seen as the consequence of a reduced cost due to the reduced number of neighbors. We discuss the general case of two thresholds $m_{\mathrm{eff}}^A$ and $m_{\mathrm{eff}}^B$ subsequently. In the following equation we interpret $x$ as the fraction $a^0$ and $1-x$ as the fraction $b^0$.

\begin{equation}
\dot{x}=\sum_{j=m_{\mathrm{eff}}}^{k_{\mathrm{eff}}} j \binom{k_{\mathrm{eff}}}{j}\left[ \lambda \left(1-x\right)^j x^{k_{\mathrm{eff}}-j}-x^j \left(1-x\right)^{k_{\mathrm{eff}}-j} \right],
\label{eq:general}
\end{equation} 
where $\lambda=a^+/b^+$. All other constant values have been neglected to increase readability and to ensure analytical tractability. Equation \eqref{eq:general} is an approximation of the time evolution of $a^0$ and $b^0$ as described by Eq.~\eqref{eq:rate}. It is taking into account the model's important features and thus providing a framework to better understand the properties of the stationary solution.

We want to find out if the function has only one fixed point. Our first observation is that the right-hand side of Eq.~\eqref{eq:general} defined as $f(x;m_{\mathrm{eff}},k_{\mathrm{eff}},\lambda)$ approaches $f(x;m_{\mathrm{eff}},k_{\mathrm{eff}},\lambda)=k_{\mathrm{eff}} \lambda$ for $x\rightarrow 0$ and $f(x;m_{\mathrm{eff}},k_{\mathrm{eff}},\lambda)= -k_{\mathrm{eff}}$ for $x\rightarrow 1$. If the function $f(x;m_{\mathrm{eff}},k_{\mathrm{eff}},\lambda)$ monotonically decreased in $x$ in the interval $(0,1)$, there would be only a unique fixed point. For $m_{\mathrm{eff}}=0,1$, the derivative with respect to $x$ is given by $f^\prime(x;m_{\mathrm{eff}},k_{\mathrm{eff}},\lambda)=-k_{\mathrm{eff}} (1+\lambda)$ and for $m_{\mathrm{eff}}=k_{\mathrm{eff}}$ the derivative is $f^\prime(x;m_{\mathrm{eff}},k_{\mathrm{eff}},\lambda)=-k_{\mathrm{eff}}^2 \left[x^{k_{\mathrm{eff}}-1} + \lambda \left(1-x\right)^{k_{\mathrm{eff}}-1}\right]$. In both cases, the derivative indicates a monotonically decreasing function and thus the existence of a unique fixed point.

For a general threshold $m_{\mathrm{eff}}$ the derivative can be obtained using \emph{Mathematica} \cite{Mathematica},
\begin{align}
\begin{split}
f^\prime(x;m_{\mathrm{eff}},k_{\mathrm{eff}},\lambda)=&-\left(\Gamma(k_{\mathrm{eff}}+1-m_{\mathrm{eff}}) \Gamma(m_{\mathrm{eff}}+1)\right)^{-1} k_{\mathrm{eff}} ((1-x) x)^{-m_{\mathrm{eff}}-1}\Gamma(k_{\mathrm{eff}}) \\
 & \left[(1 - x)^{2 m_{\mathrm{eff}}} x^{k_{\mathrm{eff}}+1} \lambda \left(m_{\mathrm{eff}} ( m_{\mathrm{eff}} -1 + x) + k_{\mathrm{eff}}(1- 
             x) \, _2F_1\left(1,m-k_{\mathrm{eff}};\,m_{\mathrm{eff}}+1;\, \frac{x-1}{x} \right)\right) \right.+ \\
             & (1 - x)^{k_{\mathrm{eff}}+1} x^{2 m_{\mathrm{eff}}} \left( m_{\mathrm{eff}} (m_{\mathrm{eff}} - x) + 
          \left. k_{\mathrm{eff}}  x  \, _2F_1\left(1,m_{\mathrm{eff}}-k_{\mathrm{eff}};\,m_{\mathrm{eff}}+1;\, \frac{x}{x-1} \right) \right) \right].
\end{split}
\end{align}
Since $x$ takes values in the interval $(0,1)$ and $m_{\mathrm{eff}} \in \left\{2, \dots, k_{\mathrm{eff}}-1\right\}$, we only have to verify if the hypergeometric functions $_2F_1$ are positive. Then $f^\prime(x;m_{\mathrm{eff}},k_{\mathrm{eff}},\lambda)$ would be negative. Due to the fact that $m_{\mathrm{eff}}-k_{\mathrm{eff}}<0$, the hypergeometric function takes the form \cite{Mathematica},
\begin{equation}
_2F_1\left(1,m_{\mathrm{eff}}-k_{\mathrm{eff}};\,m_{\mathrm{eff}}+1;\, \frac{x}{x-1}\right)=\sum_{n=0}^{k_{\mathrm{eff}}-m_{\mathrm{eff}}} (-1)^n \binom{k_{\mathrm{eff}}-m_{\mathrm{eff}}}{n} \frac{(1)_n}{(m_{\mathrm{eff}}+1)_n} \left(\frac{x}{x-1}\right)^n,
\end{equation}
where $(\cdot)_n$ denotes the Pochhammer symbol. The last term $\frac{x}{x-1}$ is smaller than 0 and thus the terms of the sum are positive, since
\begin{equation}
_2F_1\left(1,m_{\mathrm{eff}}-k_{\mathrm{eff}};\,m_{\mathrm{eff}}+1;\, \frac{x}{x-1}\right)=\sum_{n=0}^{k_{\mathrm{eff}}-m_{\mathrm{eff}}} \binom{k_{\mathrm{eff}}-m_{\mathrm{eff}}}{n} \frac{(1)_n}{(m_{\mathrm{eff}}+1)_n} \left(\frac{x}{1-x}\right)^n.
\end{equation}
The same argument applies to $_2F_1\left(1,m_{\mathrm{eff}}-k_{\mathrm{eff}};\,m_{\mathrm{eff}}+1;\, \frac{x-1}{x}\right)$. Therefore, $f^\prime(x;m_{\mathrm{eff}},k_{\mathrm{eff}},\lambda)<0$ and the basic model as defined by Eq.~\eqref{eq:general} has one unique stationary state.
\begin{figure}
\begin{minipage}{0.45\textwidth}
\centering
\includegraphics[width=\textwidth]{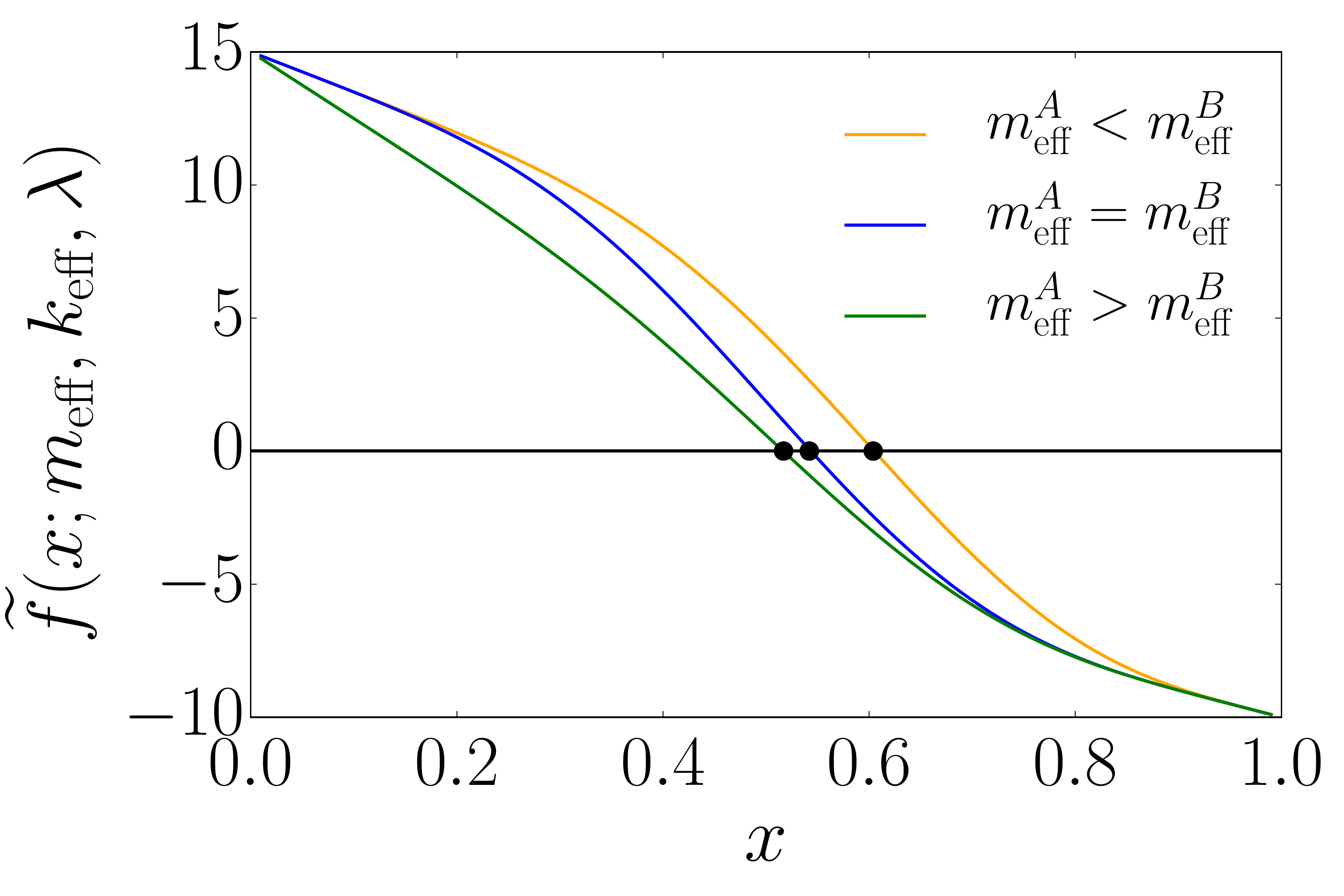}
\end{minipage}
\begin{minipage}{0.45\textwidth}
\centering
\includegraphics[width=\textwidth]{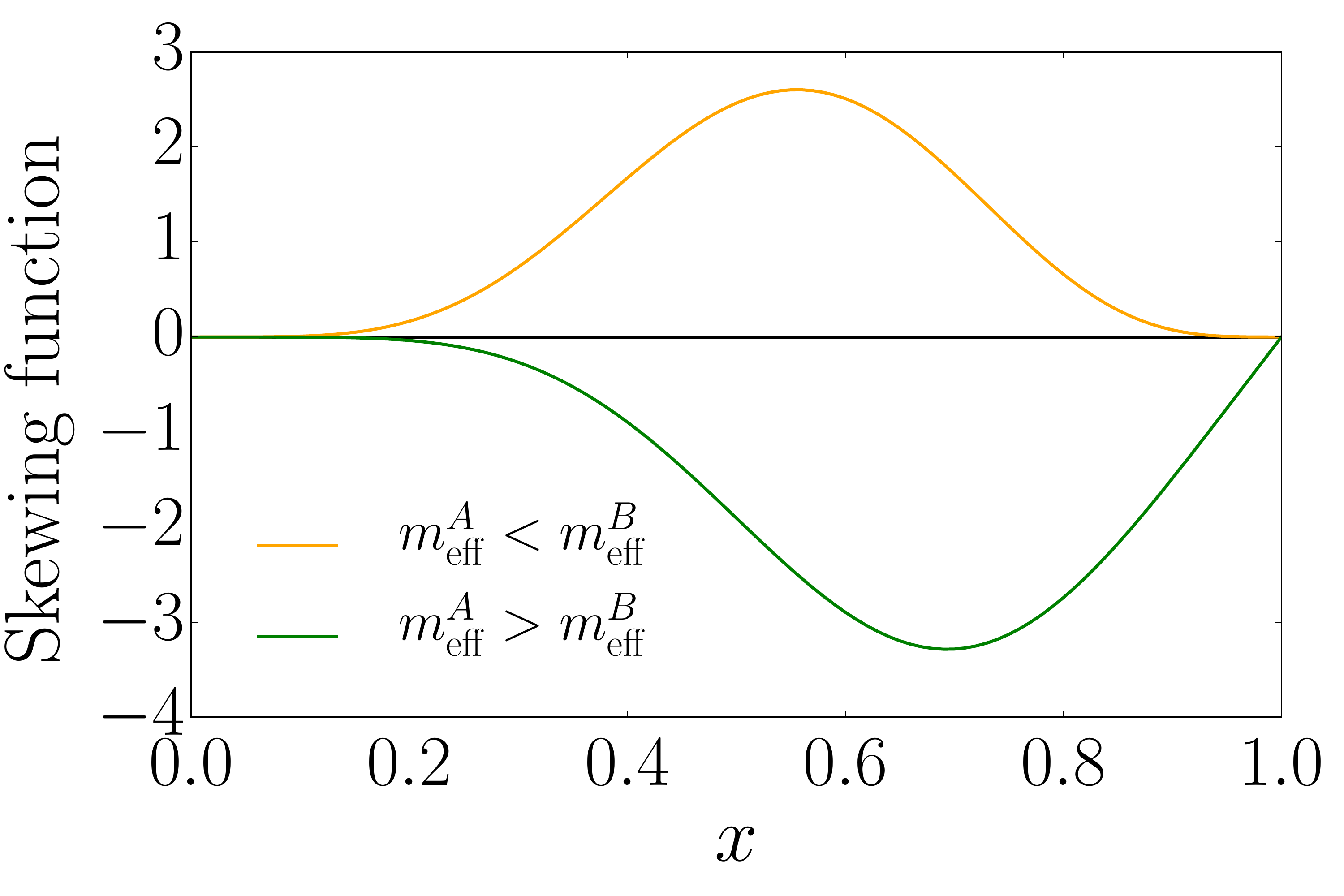}
\end{minipage}
  \caption{\textbf{General phase portrait and skewing functions.} The left panel shows the three possible cases of the campaign model's phase portrait where $\widetilde{f}\left(x;m^A_{\mathrm{eff}},m^B_{\mathrm{eff}},k_{\mathrm{eff}},\lambda\right)$ defines the right-hand side of Eq.~\eqref{eq:general2}. We set $k_{\mathrm{eff}}=10$, $\lambda=1.5$ and $m^A_{\mathrm{eff}}=m^B_{\mathrm{eff}}=5$ (blue solid line) whereas $m^A_{\mathrm{eff}}=5<m^B_{\mathrm{eff}}=7$ (orange solid line) and $m^A_{\mathrm{eff}}=5>m^B_{\mathrm{eff}}=0$ (green solid line). In the right panel, we define the second sums of Eqs.~\eqref{eq:general3} and \eqref{eq:general4} as the skewing functions that skew the monotonically decreasing first sums in positive ($m^A_{\mathrm{eff}}<m^B_{\mathrm{eff}}$) and negative ($m^A_{\mathrm{eff}}>m^B_{\mathrm{eff}}$) y-direction, respectively.} 
 \label{fig:skewing}
\end{figure}
We now focus on the more general case with two thresholds, i.e.~$m^A_{\mathrm{eff}}=\ceil*{\frac{c_A}{\rho_A}} (1-a^+-b^+-o)$, $m^B_{\mathrm{eff}}=\ceil*{\frac{c_B}{\rho_B}} (1-a^+-b^+-o)$ and $\lambda=\left(\rho_A a^+\right)/\left(\rho_B b^+\right)$. Thus, the resulting dynamics is described by
\begin{equation}
\dot{x}=\sum_{j=m^A_{\mathrm{eff}}}^{k_{\mathrm{eff}}} j \binom{k_{\mathrm{eff}}}{j} \lambda \left(1-x\right)^j x^{k_{\mathrm{eff}}-j}-\sum_{j=m^B_{\mathrm{eff}}}^{k_{\mathrm{eff}}} j \binom{k_{\mathrm{eff}}}{j} x^j \left(1-x\right)^{k_{\mathrm{eff}}-j}.
\label{eq:general2}
\end{equation}
Again, the right-hand side of Eq.~\eqref{eq:general2} approaches $\widetilde{f}\left(x;m^A_{\mathrm{eff}},m^B_{\mathrm{eff}},k_{\mathrm{eff}},\lambda\right)=k_{\mathrm{eff}} \lambda$ as $x \rightarrow 0$ and $\widetilde{f}\left(x;m^A_{\mathrm{eff}},m^B_{\mathrm{eff}},k_{\mathrm{eff}},\lambda\right)=-k_{\mathrm{eff}}$ as $x \rightarrow 1$. In order to show that $\widetilde{f}\left(x;m^A_{\mathrm{eff}},m^B_{\mathrm{eff}},k_{\mathrm{eff}},\lambda\right)$ decreases monotonically in $x$ in the interval $(0,1)$, we apply the previous proof by discussing the two cases: (i) $m^A_{\mathrm{eff}}<m^B_{\mathrm{eff}}$ and (ii) $m^A_{\mathrm{eff}}>m^B_{\mathrm{eff}}$. We find
\begin{equation}
\mathrm{(i)~}\dot{x}=\sum_{j=m^A_{\mathrm{eff}}}^{k_{\mathrm{eff}}} j \binom{k_{\mathrm{eff}}}{j} \left[ \lambda \left(1-x\right)^j x^{k_{\mathrm{eff}}-j} - x^j \left(1-x\right)^{k_{\mathrm{eff}}-j} \right]+\sum_{j=m^A_{\mathrm{eff}}}^{m^B_{\mathrm{eff}}-1}j \binom{k_{\mathrm{eff}}}{j} x^j \left(1-x\right)^{k_{\mathrm{eff}}-j},
\label{eq:general3}
\end{equation}
\begin{equation}
\mathrm{(ii)~}\dot{x}=\sum_{j=m^B_{\mathrm{eff}}}^{k_{\mathrm{eff}}} j \binom{k_{\mathrm{eff}}}{j} \left[ \lambda \left(1-x\right)^j x^{k_{\mathrm{eff}}-j} - x^j \left(1-x\right)^{k_{\mathrm{eff}}-j} \right]-\sum_{j=m^B_{\mathrm{eff}}}^{m^A_{\mathrm{eff}}-1}j \binom{k_{\mathrm{eff}}}{j} \lambda \left(1-x\right)^j x^{k_{\mathrm{eff}}-j}.
\label{eq:general4}
\end{equation}
The first sums in Eqs.~\eqref{eq:general3} and \eqref{eq:general4} decrease monotonically in $x$ in the interval $(0,1)$ as shown above for Eq.~\eqref{eq:general}. The second sums in Eqs.~\eqref{eq:general3} and \eqref{eq:general4} approach 0 for $x=0$ and $x=1$ and otherwise only have positive values in the interval $(0,1)$. We show two examples of these skewing functions in Fig.~\ref{fig:skewing} (right). Consequently, as shown in Fig.~\ref{fig:skewing} (left), the function defined by the first sum of Eq.~\eqref{eq:general3} gets skewed in positive y-direction without changing the fact that it decreases monotonically. In the case of Eq.~\eqref{eq:general4}, the skewing occurs in negative y-direction, cf.~Fig.~\ref{fig:skewing} (left). Thus, there is still a unique stationary fixed point even for general thresholds that occur due to different $\rho_A$, $\rho_B$ and $c_A$, $c_B$.
\subsection{Interplay between activists, convincing probabilities and cost}
\label{sec:activists}
\begin{figure}
\begin{minipage}{0.45\textwidth}
\centering
\includegraphics[width=\textwidth]{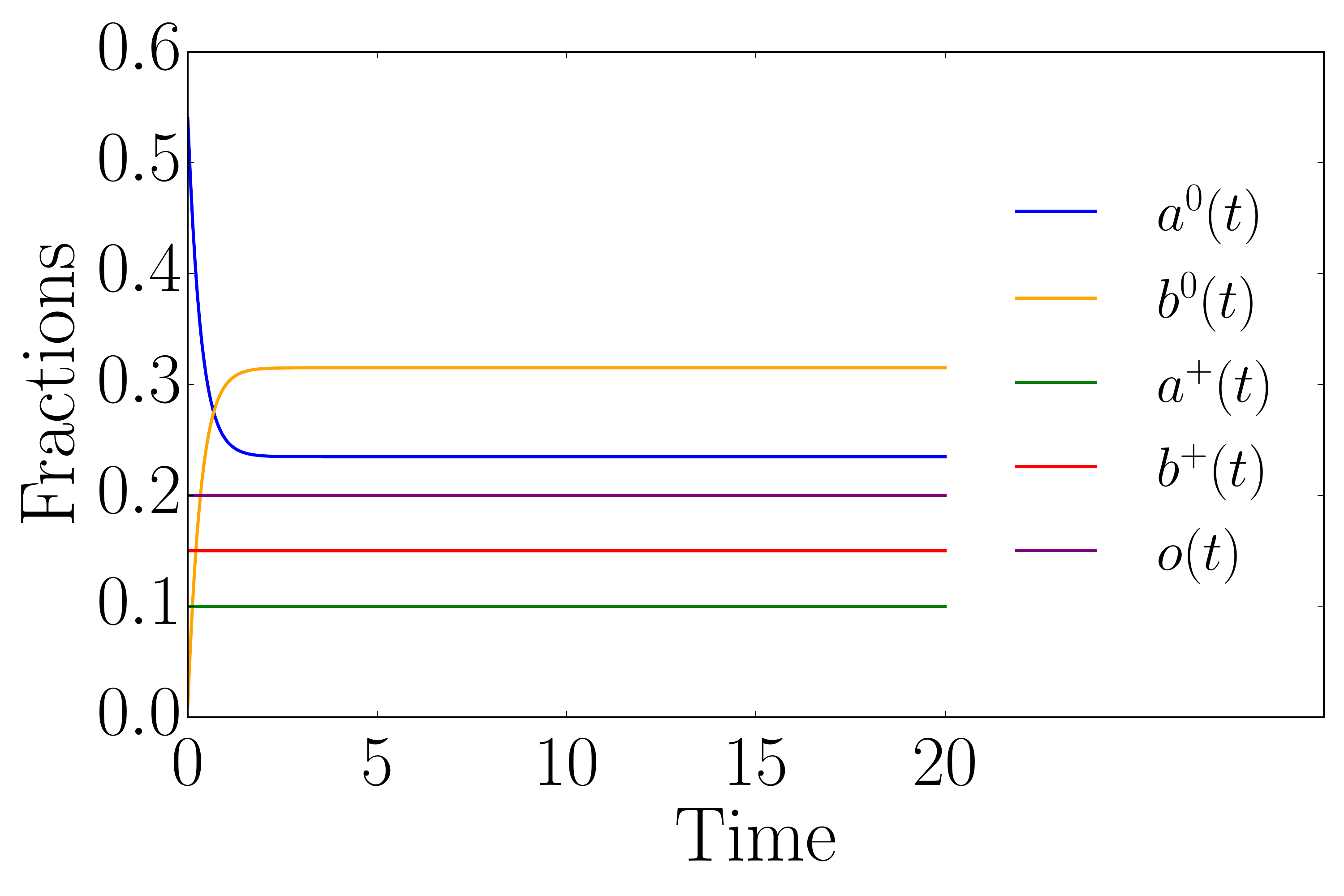}
\end{minipage}
\begin{minipage}{0.45\textwidth}
\centering
\includegraphics[width=\textwidth]{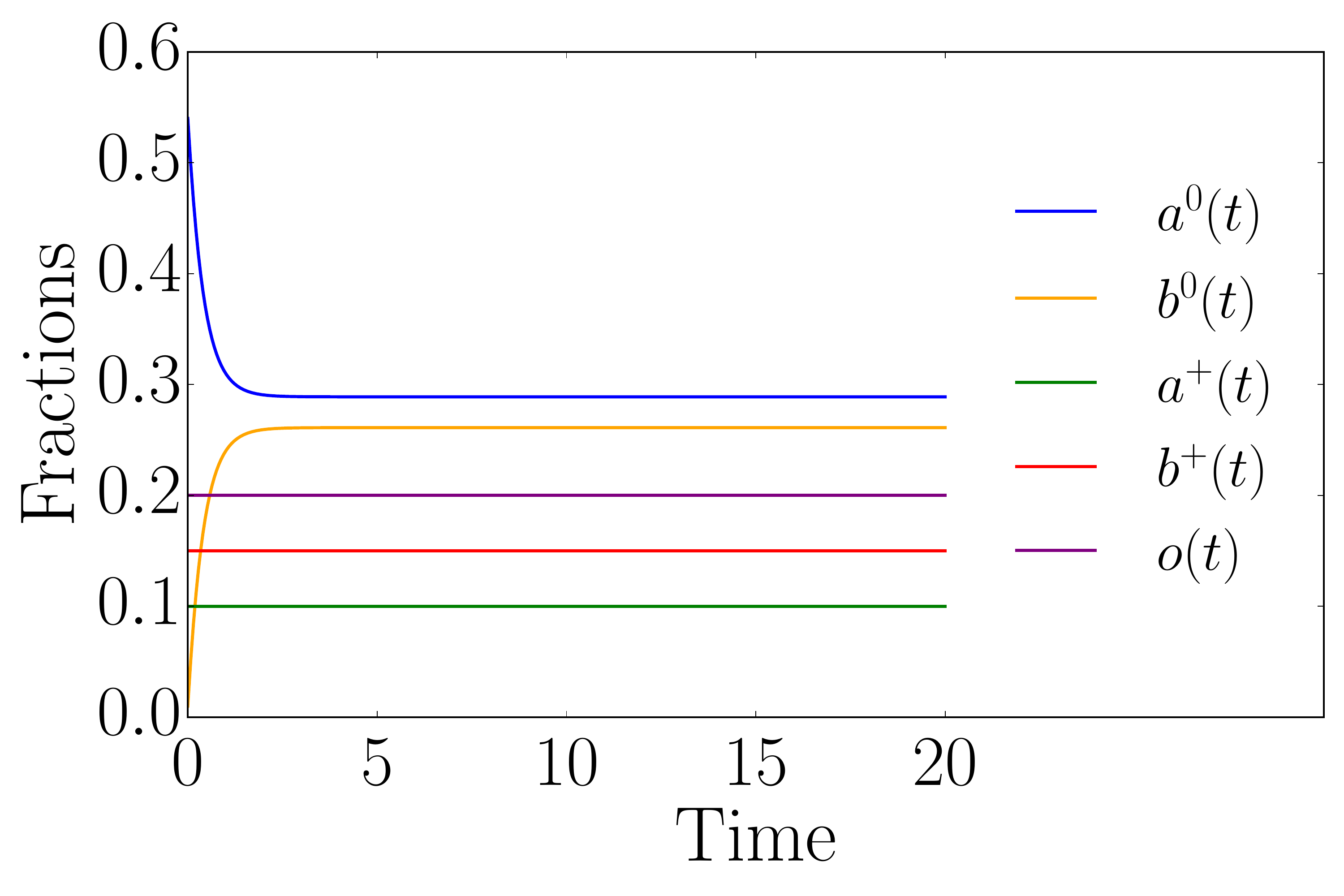}
\end{minipage}
  \caption{\textbf{Influence of activists and thresholds on the stationary states.} The time evolution of the fractions of persuadable nodes $a^0(t)$, $b^0(t)$ and activists $a^+(t)$, $b^+(t)$ as well as empty nodes $o(t)$ for two different parameter constellations are displayed. The shown curves are a numerical solution of Eq.~\eqref{eq:rate} for $k=10$ with initial conditions $a^0(0)=0.54$, $b^0(0)=0.01$, $a^+(0)=0.1$, $b^+(0)=0.15$ and $o(0)=0.2$. In the left panel, the convincing probabilities are $\rho=\rho_A=\rho_B=1.0$ and the costs are set to $c=c_A=c_B=2$. Group $B$, with the larger fraction of activists, dominates the dynamics. In the right panel, the convincing probabilities are $\rho_A=1.0$ and $\rho_B=0.8$, whereas the costs are still $c=c_A=c_B=2$. Group $A^0$ dominates due to the larger convincing probability. } 
 \label{fig:x_t_2}
\end{figure}
Intuitively, the campaign with the larger fraction of activists and the larger convincing probability should be dominant in the stationary state. As in Sec.~\ref{sec:uniqueness}, we begin the discussion by setting $\rho=\rho_A=\rho_B$, $c=c_A=c_B$. Some arguments of the model's behavior for general $\rho_A$, $\rho_B$, $c_A$, $c_B$ are given subsequently. In Fig.~\ref{fig:xdotx} we see that the stationary state of campaign group $B$ is larger than the one of group $A$, since $b^+>a^+$. Can this behavior be observed independently of the initial conditions? This would imply that only activists matter. We illustrate a solution of Eq.~\eqref{eq:rate} with $a^+(0)=0.1$, $b^+(0)=0.15$, $a^0(0)=0.54$, $b^0(0)=0.01$, $o(0)=0.2$ in Fig.~\ref{fig:x_t_2}. As in Fig.~\ref{fig:xdotx}, we again find that the group with more activists dominates the dynamics. We can also give some analytical insight into this behavior. For the moment, we assume that $\ceil*{\frac{c}{\rho}}=k$ in Eq.~\eqref{eq:rate}. Then $\dot{a}=o \rho k \left(a^+ {b^0}^k - b^+ {a^0}^k \right)$ and the corresponding stationary state reveals that $\left( \frac{a^0}{b^0}\right)^k=\left(\frac{a^+}{b^+}\right)$ in agreement with the expected influence of the activists as discussed above.

For general $\ceil*{\frac{c}{\rho}}$, we find that the stationary states are equal if $a^+=b^+$ due to the symmetric form of Eq.~\eqref{eq:rate}. We now assume that $a^+>b^+$ and focus on Eq.~\eqref{eq:general} as before.
The equation describing the stationary state $x_{\mathrm{st}}$ is given by
\begin{equation}
0=\sum_{j=m_{\mathrm{eff}}}^{k_{\mathrm{eff}}} j \binom{k_{\mathrm{eff}}}{j}\left[ \lambda \left(\frac{1-x_{\mathrm{st}}}{x_{\mathrm{st}}}\right)^j - \left(\frac{1-x_{\mathrm{st}}}{x_{\mathrm{st}}}\right)^{k_{\mathrm{eff}}-j} \right].
\label{eq:stationary}
\end{equation}
We know that there exists a unique solution $x_{\mathrm{st}}\in (0,1)$ as analytically demonstrated in Sec.~\ref{sec:uniqueness}. The solution $x_{\mathrm{st}}=0.5$ corresponds to the case where $a^+=b^+$. We now prove by contradiction that $x_{\mathrm{st}}>0.5$ if $a^+>b^+$, i.e.~$\lambda>1$. Hence, we assume that $x_{\mathrm{st}}<0.5$ if $\lambda>1$. The term $\frac{1-x_{\mathrm{st}}}{x_{\mathrm{st}}}>1$ and also $\sum_{j=m_{\mathrm{eff}}}^{k_{\mathrm{eff}}} j \binom{k_{\mathrm{eff}}}{j}\left[ \lambda \left(\frac{1-x_{\mathrm{st}}}{x_{\mathrm{st}}}\right)^j - \left(\frac{1-x_{\mathrm{st}}}{x_{\mathrm{st}}}\right)^{k_{\mathrm{eff}}-j} \right]>0$, since $\lambda>1$. The negative terms appearing for $j<k_{\mathrm{eff}}-j$ are always compensated by positive ones due to the summation over $j\in\left\{m_{\mathrm{eff}},\dots, k_{\mathrm{eff}}\right\}$, and the sum would not add up to 0. Thus, $x_{\mathrm{st}}<0.5$ cannot be a stationary state corresponding to $\lambda>1$. 
Consequently, $x>0.5$ if $a^+>b^+$. We conclude that the campaign with the larger fraction of activists exhibits a dominating stationary state if convincing probability and costs are the same.

In the case of different convincing probabilities and costs, we end up with two different effective thresholds $m^A_{\mathrm{eff}}$, $m^B_{\mathrm{eff}}$ and $\lambda=\left(\rho_A a^+\right)/\left(\rho_B b^+\right)$ as visible in Eqs.~\eqref{eq:general3} and \eqref{eq:general4}. The discussion of the latter two equations in Sec.~\ref{sec:uniqueness} revealed that additional sums just skew the monotonically decreasing right-hand sides in positive ($m^A_{\mathrm{eff}}<m^B_{\mathrm{eff}}$) and negative ($m^A_{\mathrm{eff}}>m^B_{\mathrm{eff}}$) y-direction respectively. This explains the influence of convincing probabilities and costs on the stationary states. A skewing in positive y-direction in the case of $m^A_{\mathrm{eff}}<m^B_{\mathrm{eff}}$ means that the threshold $m^A_{\mathrm{eff}}=\ceil*{\frac{c_A}{\rho_A}} (1-a^+-b^+-o)$ is smaller than $m^B_{\mathrm{eff}}$ due to smaller cost and a larger convincing probability. Then the skewing in positive y-direction means that the final stationary state of campaign group
 $A$ is getting larger, i.e.~more individuals follow opinion $A$. In the same manner, the case where $m^A_{\mathrm{eff}}>m^B_{\mathrm{eff}}$ corresponds to a growing influence of campaign group $B$. In summary, there is a complex interplay between the fractions of activists, convincing probabilities and costs.
\subsection{Further effects resulting from the influence of convincing probability and cost}
\label{sec:cost_influence}
\begin{figure}
\centering
\includegraphics[width=0.45\textwidth]{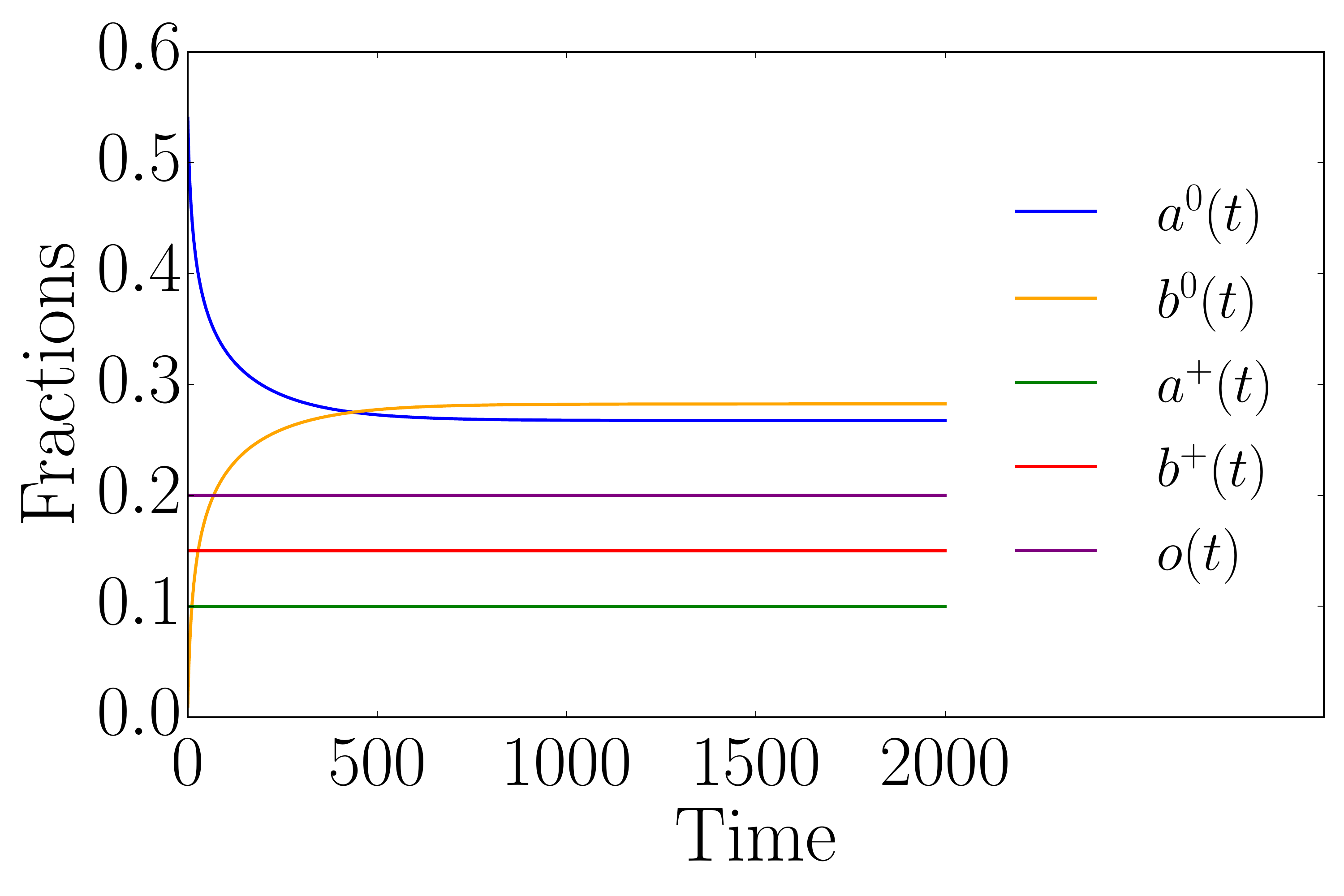}
  \caption{\textbf{Influence of costs and convincing probability.} The time-evolution of the fractions of persuadable nodes $a^0(t)$, $b^0(t)$ and activists $a^+(t)$, $b^+(t)$ as well as empty nodes $o(t)$ is displayed. The shown curves are a numerical solution of Eq.~\eqref{eq:rate} for $k=10$ with initial conditions $a^0(0)=0.54$, $b^0(0)=0.01$, $a^+(0)=0.1$, $b^+(0)=0.15$ and $o(0)=0.2$. The utility is fixed to $\rho=\rho_A=\rho_B=0.5$ and the costs are $c=c_A=c_B=4$.} 
 \label{fig:x_t_3}
\end{figure}
\begin{figure}
\begin{minipage}{0.45\textwidth}
\centering
\includegraphics[width=\textwidth]{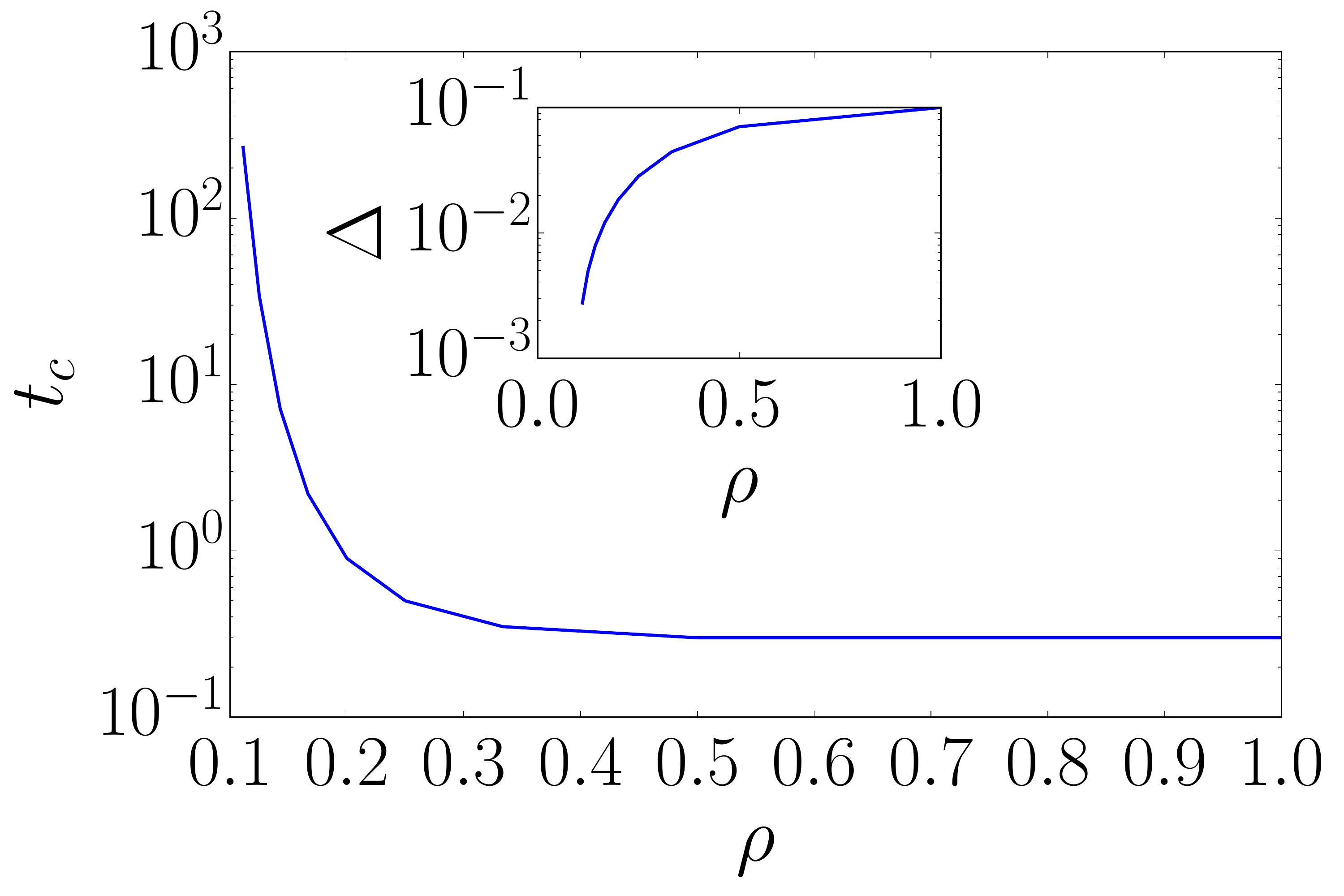}
\end{minipage}
\begin{minipage}{0.45\textwidth}
\centering
\includegraphics[width=\textwidth]{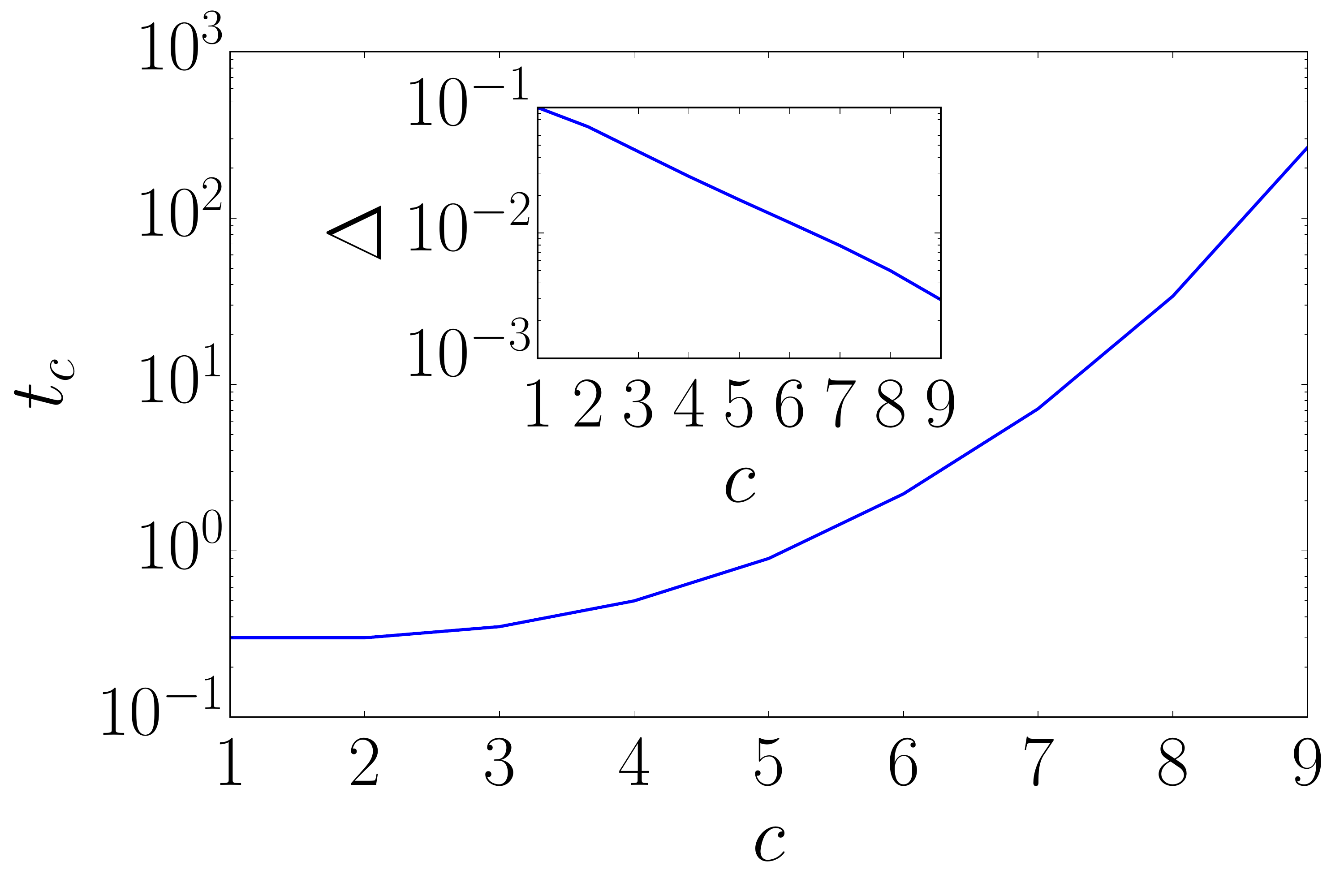}
\end{minipage}
  \caption{\textbf{Characteristic timescales and opinion gap.} The left panel show that a characteristic transient time-scale $t_c$ decreases as $\rho=\rho_A=\rho_B$ increases  The gap $\Delta=|b^0_{\mathrm{st}}-a^0_{\mathrm{st}}|$ between both stationary states, however, increases with $\rho$. The cost is set to $c=c_A=c_B=1$. From the right panel, we find that the characteristic time scale $t_c$ increases with the cost $c$, whereas the gap $\Delta$ decreases with $c$, as illustrated in the inset. The convincing probability has been fixed to $\rho=1$. We numerically solved Eq.~\eqref{eq:rate} for $k=10$, with $a^0(0)=0.54$, $b^0(0)=0.01$, $a^+(0)=0.1$, $b^+(0)=0.15$ and $o(0)=0.2$.} 
 \label{fig:t_c}
\end{figure}
In the previous Sec.~\ref{sec:activists}, we have seen that not only the fractions of activists matter but that a change in convincing probability and cost also shifts the stationary states to the advantage or disadvantage of one campaign group. There are, however, other effects originating from the influence of convincing probability and costs that shall be discussed here. To properly analyze this influence and not to deal with too many parameters, we again set $\rho=\rho_A=\rho_B$ and $c=c_A=c_B$.

We first want to discuss the influence of the parameter $\rho$ on the saturation time. The parameter $\rho$ corresponds to a time-rescaling, since the increment of Eq.~\eqref{eq:rate}, similar to $\dot{x}=\rho f(x)$, comes with a prefactor $\rho$. Smaller values of $\rho$ would intuitively lead to longer transient times. Furthermore, $\rho$ also enters the threshold condition of Eq.~\eqref{eq:rate} whose summation starts at $\ceil*{\frac{c}{\rho}}$ . We remember that no dynamics occurs if $\ceil*{\frac{c}{\rho}}>k$ since this would imply that the number of persuadable nodes has to be greater than the actual degree of the network. Also in terms of the threshold condition, we would expect longer transient times for smaller values of $\rho$. Larger threshold values correspond to the situation where activists require a greater number of persuadable nodes in order to perform an action, cf.~Fig.~\ref{fig:x_t_3}. We illustrate the dependence of a characteristic transient time-scale $t_c$ on $\rho$ in Fig.~\ref{fig:t_c} (left). In agreement with the qualitative arguments given above, we find that $t_c$ decreases as $\rho$ increases. Moreover, the inset in Fig.~\ref{fig:t_c} (left) shows that the gap $\Delta=|b^0_{\mathrm{st}}-a^0_{\mathrm{st}}|$ between the stationary states increases with $\rho$.

Similar to the effect of a small convincing probability $\rho$, a large cost parameter $c$ would intuitively lead to longer transients as shown in Fig.~\ref{fig:x_t_3}. Not all neighborhoods fulfill the utility condition in the case of a large cost and consequently it takes more time to reach the equilibrium. We also illustrate this behavior in Fig.~\ref{fig:t_c} (right). As the cost parameter $c$ increases, the characteristic time-scale $t_c$ is getting larger. Moreover, we see in the inset of Fig.~\ref{fig:t_c} (right) that the gap $\Delta$ between the two stationary opinions shrinks as the cost increases.

The following example illustrates an interesting effect occurring as a consequence of the shrinking gap for small values of $\rho$ or large values of $c$. First, we compare Figs.~\ref{fig:x_t_2} (left) and S\ref{fig:x_t_3}. We find that $a^0_{\mathrm{st}}\approx 0.235$ and $b^0_{\mathrm{st}}\approx 0.315$ in Fig.~\ref{fig:x_t_2} (left), where $c=2$ and $\rho=1$. These values correspond to a gap of $\Delta = 0.08$. In Fig.~\ref{fig:x_t_3} with $c=4$ and $\rho=0.5$ we, however, find a smaller gap of $\Delta = 0.014$, since $a^0_{\mathrm{st}}\approx 0.268$ and $b^0_{\mathrm{st}}\approx 0.282$. Taking into account the fractions of activists in both cases, i.e.~$a^+=0.1$ and $b^+=0.15$, we conclude that 58 \% are in favor of opinion B in the first case whereas this value is reduced to 54 \% in the second case. In addition, we could think about a change of the majority structure as a consequence of the influence of $\rho$ and $c$. We assume another compartment of inactive individuals with a fixed opinion $a^-$ and $b^-$. They are not relevant for the dynamics at all but influence the majority structure. We now set $a^-=0.1$ and $b^-=0.02$, thereby implicitly reducing the value of $o$. As a consequence, 53 \% of the individuals are in favor of opinion $B$ in the first case but in the second case, this campaign group loses its majority with 49 \%.
%
%
%
%
%
%
%
%
%
%
%
%
%
\end{document}